\documentclass[pra,twocolumn,preprintnumbers,amsmath,amssymb,superscriptaddress,longbibliography]{revtex4-1}
\usepackage{graphicx}
\usepackage{dsfont}
\usepackage{natbib}
\usepackage[squaren]{SIunits}

\allowdisplaybreaks

\newcommand{\aVec}
					{ a }

\usepackage{color}

\usepackage{ulem} 

\begin{document}

\title{Influence of nuclear spin polarization on spin echo signal of NV center qubit}
\author{Damian Kwiatkowski}
\affiliation{Institute of Physics, Polish Academy of Sciences, al.~Lotnik{\'o}w 32/46, PL 02-668 Warsaw, Poland}
\author{Piotr Sza\'{n}kowski}
\affiliation{Institute of Physics, Polish Academy of Sciences, al.~Lotnik{\'o}w 32/46, PL 02-668 Warsaw, Poland}
\author{{\L}ukasz Cywi\'{n}ski}\email{lcyw@ifpan.edu.pl}
\affiliation{Institute of Physics, Polish Academy of Sciences, al.~Lotnik{\'o}w 32/46, PL 02-668 Warsaw, Poland}

\begin{abstract}
We consider the spin echo dynamics of a nitrogen-vacancy center qubit based the $S\!= \! 1$ ground state spin manifold, caused by a dynamically polarized nuclear environment. We show that the echo signal acquires then a nontrivially time-dependent phase shift. This effect should be observable for polarization $\approx \! 0.5$ of nuclei within $\sim \! 1$ nm from the qubit, and for the NV center initialized in a superposition of $m\! = \! 0$ and either $m\! =\! 1$ or $m\! =\! -1$ states.
This phase shift is much smaller when the NV center is prepared in a superposition of $m\! = \! 1$ and $m\! =\! -1$ states, i.e.~when the qubit couples to the spin environment in a way analogous to that of spin-$1/2$. For nuclear environment devoid of spins strongly coupled to the qubit, the phase shift is well described within Gaussian approximation, which provides an explanation for the dependence of the shift magnitude on the choice of states on which the qubit is based, and makes it clear that its presence is related to the linear response of the environment perturbed by an evolving qubit. Consequently, its observation signifies the presence environment-mediated self-interaction of the qubit, and hence, it invalidates the notion that the nuclear environment acts as a source of external noise driving the qubit. We also show how a careful comparison of the echo signal from qubits based on $m\! = \! 0,1$ and $m\! =\! \pm 1$ manifolds, can distinguish between effectively Gaussian and non-Gaussian environment.
\end{abstract}

\date{\today}
		
\maketitle

\section{Introduction}
When a qubit undergoes pure dephasing due to interaction with its environment, while being subjected to an appropriate dynamical decoupling control sequence, the measurement of its coherence gives access to an abundance of information about the dynamics of environmental degrees of freedom \cite{Degen_RMP17,Szankowski_JPCM17}. The process of extraction of this information is particularly straightforward when dephasing can be described within Gaussian approximation \cite{Cywinski_PRB08,Biercuk_JPB11,Degen_RMP17,Szankowski_JPCM17}, in which the environmental fluctuations affecting the qubit's phase are fully characterized by the spectral density---the Fourier transform of the relevant autocorrelation function of environmental variables. Such reconstruction protocols, connecting the observed time-dependence of coherence to putative spectrum of noise, have been widely used during the last decade for various types of qubits \cite{Degen_RMP17,Szankowski_JPCM17}.

The most commonly encountered qubit-environment coupling leading to pure dephasing is of the form
\begin{equation}
\hat{H}_{\mathrm{int}} = \frac{\lambda}{2} \hat{\sigma}_{z}\otimes \hat{V},
\end{equation}
where $\lambda$ is a dimensionless parameter and eigenstates of $\hat{\sigma}_z$ = $|{\uparrow}\rangle\langle{\uparrow}|-|{\downarrow}\rangle\langle{\downarrow}|$ is the $z$ Pauli operator. In an appropriate rotating frame, the off-diagonal element of the qubit's reduced density matrix at time $t$ is given in the Gaussian approximation by
\begin{align}
 \label{eq:W}
\rho_{{\uparrow}{\downarrow}}(t) & = \langle{\uparrow}| \,\mathrm{Tr}_{E}( \hat{U}(t) \hat{\rho}_{QE}(0) \hat{U}^{\dagger}(t) )\,|{\downarrow}\rangle
	= \rho_{{\uparrow}{\downarrow}}(0) e^{-\lambda^2 \chi(t)},
\end{align}
where  $\mathrm{Tr}_{E}$ is partial trace with respect to the environment, $\hat{U}(t)$ is the evolution operator of the qubit--environment composite system that includes the dynamical control exerted on the qubit, and the real-valued and non-negative $\chi(t)$ that fully describes the evolution is called the {\it attenuation function}.

The discussion presented here is motivated by Ref.~\cite{Paz_PRA17}, in which the qubit dephasing due to interaction with an environment composed of non-interacting bosons was considered. Crucially, a more general form of qubit-environment coupling leading to pure dephasing was considered there: 
\begin{equation}
\hat{H}_{\mathrm{int}} = \frac{1}{2}\lambda(\eta\hat{\mathds{1}} + \hat\sigma_z)\otimes\hat V . \label{eq:biased}
\end{equation}
For $\eta \neq 0$ ($\eta = 0$) we will refer to it as a {\it biased} ({\it unbiased}) coupling. It was observed there, that for biased coupling, the coherence differs form Eq.~\eqref{eq:W} by a phase factor,
\begin{equation}
\rho_{{\uparrow}{\downarrow}}(t) \! =\!\rho_{{\uparrow}{\downarrow}}(0) e^{-\lambda^2 \chi(t)} e^{-i\eta\lambda^2 \Phi(t)}. \label{eq:Wasym}
\end{equation}
The additional {\it bias-induced} phase shift $\Phi(t)$ has a nontrivial time-dependence determined by the dynamical properties of the environment \cite{Paz_PRA17}, and it disappears when the state of the environment is maximally mixed, $\hat\rho_E \propto \hat{\mathds{1}}$. Hence, in order to observe this shift, the inverse temperature $\beta$ has to be finite, in the case of thermal equilibrium. Alternatively, one could intentionally drive the environment into a non-equilibrium state, and steer its state away from the maximal mixture. This approach is especially relevant for this paper, as we will focus here on the case of nuclear spin environment subjected to the dynamic nuclear polarization (DNP).

As discussed extensively in \cite{Paz_PRA17}, when one employs an approximation where the environment-induced qubit dynamics are simulated with an external field (a noise), the phase shift $\Phi(t)$ cannot appear, and the bias in coupling has no impact on the evolution. This is simply because any shift of the energy scale (even a time dependent one) does not affect the energy splitting between the qubit levels, and thus, has no influence over coherence. In other words, the observation of nonzero $\Phi(t)$ under dynamical decoupling in Gaussian dephasing regime, signifies that the influence of the environment {\it cannot} be treated as an external disturbance in form of a noise field.

The biased coupling discussed above, arises in natural way for a qubit that is based on $m\! =\! 0$ and $1$ levels of spin $S\! =\! 1$  system. This applies to widely investigated spin qubits based on nitrogen-vacancy (NV) center in diamond \cite{Dobrovitski_ARCMP13,Rondin_RPP14}. A typical environment of such a qubit consist of nuclear spins of ${}^{13}\mathrm{C}$ \cite{Coish_PSSB09,Cywinski_APPA11,Chekhovich_NM13} and the decoherence process they induce is not necessarily Gaussian. However, there are regimes of environment sizes, timescales, and qubit--environment coupling strengths in which the Gaussian approximation holds \cite{Szankowski_JPCM17,Kwiatkowski_PRB18}. In this paper, we consider the spin echo dynamics of such a system. We investigate the effects of DNP applied to the environmental nuclei; a method successfully implemented in various experimental settings  \cite{London2013,Fischer2013a,Pagliero2018,Wunderlich2017,Alvarez2015,King2015,Scheuer2017,Poggiali2017,Hovav2018}. We calculate the exact echo signal at magnetic field of $\unit{15.6}{\milli\tesla}$, and analyze the bias-induced phase shift of the echoed coherence using the method of cluster-correlation expansion (CCE) \cite{Yang_CCE_PRB08,Zhao_PRB12,Yang_RPP17,Kwiatkowski_PRB18} that allows for a well-controlled (essentially exact at timescales relevant for results presented here) account of inter-nuclear interactions. We show that the phase shift should be easily observable in experiments: polarization degree of $0.5$ for nuclei located up to $\approx \unit{2}{\nano\meter}$ from the center results in a phase that is at least a fraction of $\pi$. While the Gaussian approximation {\it quantitatively} describes the NV center decoherence when there are no nuclear spins in $\approx \unit{1}{\nano\meter}$ radius from the qubit, the most relevant for us here {\it qualitative} feature of the Gaussian result---the appearance of sizable polarization-induced phase for a biased qubit---apply even in the cases when the Gaussian approximation is not valid. We discuss a couple of ways in which the presence of non-Gaussian character of environmental influence on the qubit can be ascertained by comparing echo signals in the biased and unbiased case.

The paper is organized in the following way. In Section \ref{sec:Gaussian} we revisit the general theory of pure dephasing and derive a physically transparent expression for the phase $\Phi(t)$ that arises for biased qubit--environment coupling. This result was given first in \cite{Paz_PRA17} for a general multi-qubit case, and here for completeness we present a simpler derivation in a single qubit case, which is sufficient for our purposes. Then in Section \ref{sec:NV} we revisit the most practical approach to calculation of electron spin qubit decoherence caused by a nuclear environment, the CCE method, and discuss its generalization to the case of finite DNP. Subsequently, we discuss in Section \ref{sec:WCvsG} the qualitative conditions that allow for using Gaussian approximation \cite{Szankowski_JPCM17} when dealing with an environment that can be naturally decomposed into weakly interacting subsystems (e.g.~a spin bath with weak intra-bath interactions), and we explain the distinction between dephasing calculated in the weak-coupling approximation (i.e.~when each of the subsystems is a source of a very small perturbation to qubit's coherence) and in the Gaussian approximation. This sets the foundation for subsequent discussion of $\Phi(t)$ generated by a nuclear spin environment. In Section \ref{sec:results} we demonstrate the numerical results for spin echo dynamics at low magnetic fields ($B_0 \! = \! \unit{15.6}{\milli\tesla}$), for which the most relevant environmental dynamical process leading to the decay is the Larmor precession of single nuclear spins. We compare the results of the full CCE calculations with Gaussian approximation results, and show that the latter approximation is applicable for environments that do not contain spins that are close to the qubit. The phase shift of the echo signal is shown there to appear whenever there is finite DNP, and the qubit is subjected to the biased coupling.
Comparison between decoherence of the qubit based on $m\! =\! 0$ and $m \! = \! 1$ levels of the ground-state spin $S\! =\! 1$ manifold of the NV center (biased), with the qubit based on $m\! =\! \pm 1$ levels (unbiased) is also given there. We discuss how by comparing the coherence signal obtained for these two kinds of qubits one can detect the non-Gaussian features of environmental fluctuations leading to qubit decoherence. In the concluding Section \ref{sec:conclusions} we  summarize the results and give examples of other qubits, for which analogous effects could be observed. 

\section{Pure dephasing of a qubit in Gaussian approximation} \label{sec:Gaussian}
Consider a qubit--environment complex where the parties are coupled through interaction Hamiltonian \eqref{eq:biased}, that commutes with qubit's free Hamiltonian $\hat H_Q \propto \hat\sigma_z$. In such a case, the qubit undergoes pure dephasing, and only the off-diagonal elements of its reduced density matrix change during the evolution,
\begin{align}
\nonumber
\hat\rho_Q(t) &= \left(
	\begin{array}{cc}
	\rho_{{\uparrow}{\uparrow}}(0) & \rho_{{\uparrow}{\downarrow}}(t)\\
	\rho_{{\uparrow}{\downarrow}}^*(t) & \rho_{{\downarrow}{\downarrow}}(0)\\
	\end{array}
\right)\\
&\equiv\left(
	\begin{array}{cc}
	\rho_{{\uparrow}{\uparrow}}(0) & \rho_{{\uparrow}{\downarrow}}(0) W(t) \\
	\rho_{{\uparrow}{\downarrow}}^*(0) W^*(t) & \rho_{{\downarrow}{\downarrow}}(0)\\
	\end{array}\right).
\end{align}
The so-called {\it coherence function} $W(t)$ encapsulates all the changes in qubit's state caused by the interaction with the environment.

In addition, assume that during its evolution, the qubit can be influenced by an application of external control fields. The control scheme we focus on in particular is of the {\it dynamical decoupling} \cite{Degen_RMP17,Szankowski_JPCM17,Biercuk_JPB11,Viola_PRA98,Suter_RMP16} type: the qubit is subjected to the sequence of specifically timed pulses of transverse field that cause an effectively instantaneous $\pi$-rotations of its Bloch vector (spin flips). For example, the spin echo (which is the main focus of this paper) is a special case of such a sequence, where a single pulse is applied at the midpoint of the evolution. In general, the effects of dynamical decoupling control defined by the sequence of pulse application timings, $\{ \tau_0 = 0, \tau_1,\tau_2,\ldots,\tau_{n-1}, \tau_n = t\}$ (here, $t$ is the total evolution duration), can be conveniently parametrized with the so-called {\it time-domain filter function} $f(\tau)$:  
\begin{align}
\label{eq:Wdef}
&W(t) = \frac{
	\mathrm{Tr}\left(\hat\sigma_{-}\otimes\hat{\mathds{1}}\left(\hat U(t|f)\hat\rho_Q(0)\otimes\hat\rho_E(0)\hat U^\dagger(t|f)\right)\right)
	}
	{\rho_{{\uparrow}{\downarrow}}(0)}  
\end{align}
with the unitary evolution operator conditioned by the choice of pulse sequence given by
\begin{align}
&\hat U(t|f) = \mathcal{T}e^{-\frac{i\lambda}{2}\int_0^t d\tau(\eta\hat{\mathds{1}} + \hat\sigma_z f(\tau))\otimes\hat V(\tau)},   \label{eq:Udef}
\end{align}
where $\hat V(\tau) = {\exp}(i\tau \hat H_E)\hat V{\exp}(-i\tau \hat H_E)$ is the interaction picture of the environmental operator that couples to the qubit. See Appendix \ref{sec:appendix:filterfunc} for detailed derivation of this result and the explicit definition of filter $f(\tau)$. In the case of spin echo, the filter has a very simple form
\begin{align}
f_\mathrm{echo}(\tau) = \Theta(\tau)\Theta(t/2-\tau) - \Theta(\tau - t/2)\Theta(t-\tau).
\end{align}
We also assume that the applied pulse sequences are always chosen so that $\int_0^t f(\tau)d\tau = 0$; this eliminates the contribution from $\hat H_\mathrm{Q}$ in Eq.~\eqref{eq:Wdef}.

Defining the symbol $\langle \bullet\rangle = \mathrm{Tr}_E[\bullet \hat \rho_E(0)]$ to denote the partial trace over environmental degrees of freedom, the coherence function can be written in a standard form of averaged ordered exponential, that can in turn can be expressed in terms of the {\it cumulant series} \cite{Szankowski_JPCM17},
\begin{align}
\label{eq:cumulant_exp}
W(t) = \mathrm{Tr}\langle \hat U^\dagger(t|f)\hat\sigma_{-}\hat U(t|f)\hat{\sigma}_{+} \rangle = {\exp}\left(\sum_{n=0}^\infty \lambda^n \kappa_n(t|f)\right),
\end{align}
where $\kappa_n(t|f)$ is a $n$th order cumulant. To obtain the explicit form of cumulants, one can expand both sides of the above relation and compare the terms of the respective orders in $\lambda$. We will do exactly that, starting with the left side of the equality
\begin{widetext}
\begin{align}
\nonumber
W(t) &= 
	\left\langle
	\left(
	1 + \frac{i\lambda}{2} \int_0^t d \tau_1 (\eta + f(\tau_1) )\hat V(\tau_1)
	-\frac{\lambda^2}{4}\int_0^t d \tau_1\int_0^{\tau_1} d\tau_2 (\eta+f(\tau_1))\hat V(\tau_1)(\eta + f(\tau_2))\hat V(\tau_2)
	\right)
	\right.\\
&\phantom{=}
	\times\left.\left(1 - \frac{i\lambda}{2}\int_0^t d\tau_2(\eta - f(\tau_2))\hat V(\tau_2)
	-\frac{\lambda^2}{4}\int_0^t d \tau_1\int_0^{\tau_1} d\tau_2 (\eta-f(\tau_2))\hat V(\tau_2)(\eta - f(\tau_1))\hat V(\tau_1)\right)
	\right\rangle + O(\lambda^3) \,\, .
\end{align}
\end{widetext}
Assuming $\langle \hat V(\tau)\rangle = 0$, we express the double, equal-time integrals as the time-ordered ones, $\int_0^t d\tau_1\int_0^t d\tau_2 = \int_0^td\tau_1\int_0^{\tau_1}d\tau_2 + \int_0^t d\tau_2\int_0^{\tau_2}d\tau_1$ which allows us to rewrite the above expression in the terms of averaged commutator and anticommutator of the coupling operator,
\begin{align}
\nonumber
&W(t) = 1 - \frac{\lambda^2}{2}\!\int_0^t\!\!\! d\tau_1\!\int_0^{\tau_1}\!\!\!\!\! d\tau_2
	 \,f(\tau_1)f(\tau_2)\mathrm{Re}\langle \{\hat V(\tau_1),\hat V(\tau_2)\}\rangle\\
\label{eq:2nd_order_W}
&\phantom{W}
	-\frac{i\lambda^2}2 \int_0^t \!\!\!d\tau_1\!\int_0^{\tau_1}\!\!\!\!\!d\tau_2\, \eta f(\tau_2)\mathrm{Im}\langle [\hat V(\tau_1),\hat V(\tau_2) ]\rangle  
	+ O(\lambda^3),
\end{align}
where we utilized the symmetries of commutator $\langle [\hat V(\tau_1),\hat V(\tau_2) ] \rangle^* = - \langle [\hat V(\tau_1),\hat V(\tau_2)]\rangle$ (it is purely imaginary), and of anticommutator $\langle \{ \hat V(\tau_1),\hat V(\tau_2)\}\rangle^* = \langle \{ \hat V(\tau_1),\hat V(\tau_2)\}\rangle$ (it is purely real).

The Gaussian approximation to qubit's dynamics terminates the cumulant series on the second term, therefore, we can write the right hand side of \eqref{eq:Wdef} as
\begin{align}
W(t) &= e^{-\lambda^2\kappa_2(t|f)} = e^{-\lambda^2\chi(t) - i \lambda^2 \Phi(t)}  \nonumber\\
\label{eq:W_Gaussian}
&=1 - \lambda^2\chi(t) - i \lambda^2\Phi(t) + O(\lambda^3).
\end{align}
Comparing Eqs.~\eqref{eq:2nd_order_W} and \eqref{eq:W_Gaussian} we identify the real and imaginary parts of \eqref{eq:2nd_order_W} with the attenuation function and the bias-induced phase shift, respectively
\begin{align}
\label{eq:chi}
&\chi(t) = \frac{1}{2}\int_0^t \!\!\!d\tau_1\!\int_0^{\tau_1}\!\!\!\!\!d\tau_2 f(\tau_1)f(\tau_2)
	\mathrm{Re}\langle \{\hat V(\tau_1),\hat V(\tau_2)\}\rangle, \\
\label{eq:Phi} 
&\Phi(t) = \frac{\eta}{2}\int_0^t \!\!\!d\tau_1\!\int_0^{\tau_1}\!\!\!\!\!d\tau_2 \,f(\tau_2)
	\mathrm{Im}\langle [\hat V(\tau_1),\hat V(\tau_2) ] \rangle .
\end{align}
These equations simply restate the general result of Ref.~\cite{Paz_PRA17} in a single-qubit setting.

If we rewrite the integrand defining $\Phi(t)$ as
\begin{align}
\nonumber
\Phi(t) &= \eta \int_0^t \mathrm{d}\tau_1 \int_0^t \mathrm{d}\tau_2 \,f(\tau_2)\\
\nonumber
&\phantom{=}\times\left(\frac{1}{2}\Theta(\tau_1-\tau_2)\mathrm{Im}\langle [\hat V(\tau_1 - \tau_2),\hat V(0) ] \rangle\right)\\
&\equiv \eta \int_0^t \mathrm{d}\tau_1 \int_0^t \mathrm{d}\tau_2  \,f(\tau_2) K(\tau_1-\tau_2),
\end{align} 
then we recognize in $K(\tau)$ the Green-Kubo {\it susceptibility}. This quantity defines the linear response of environmental observable $\hat V$ to external stimulus that probes the system by coupling to the same observable. Therefore, the following physical interpretation can be given for the origin of the phase shift. The qubit influences the environment by coupling through variable $\hat V$. In turn, this disturbance modifies the ``noise'' generated by the environment, which drives the dephasing of the qubit. In a sense, the phase shift can be understood as a self-interaction of the qubit mediated by the environment.

Similarly, the anticommutator in the attenuation function can be also be expressed in the terms of fundamental physical quantity,
\begin{align}
\nonumber
\chi(t) &= \frac{1}{2}\int_0^t d\tau_1\int_0^t d\tau_2\, f(\tau_1)f(\tau_2)\\
\nonumber
&\phantom{=}\times\left(\frac{1}{2}\mathrm{Re}\langle\{\hat V(\tau_1-\tau_2),\hat V(0)\}\rangle\right)\\
&\equiv \frac{1}{2}\int_0^t \!\!\!d\tau_1 d\tau_2\, f(\tau_1)f(\tau_2) C(\tau_1-\tau_2),
\end{align}
where $C(\tau)$ is the {\it autocorrelation function} that describes the natural temporal fluctuations of environmental observable $\hat V$. The two quantities are deeply related; specifically, for a system that exhibits detailed balance---such as environment in thermal equilibrium state $\hat\rho_E = e^{-\beta \hat H_E}/\mathrm{Tr}(e^{-\beta \hat H_E})$---the Fourier transforms of correlation function (the spectral density), 
\begin{align}
S(\omega) &= \frac{1}{2}\int_{-\infty}^\infty e^{-i \omega \tau} \,\mathrm{Re}\langle \{\hat V(\tau),\hat V\}\rangle\, d\tau
\end{align}
and of susceptibility,
\begin{align}
K(\omega) &= \frac{1}{2}\int_{-\infty}^\infty e^{-i \omega \tau}\Theta(\tau)\mathrm{Im}\langle [\hat V(\tau),\hat V]\rangle\, d\tau
\end{align}
satisfy the {\it fluctuation-dissipation theorem},
\begin{align}
S(\omega) = \mathrm{coth}\left(\frac{\beta\omega}{2}\right) \mathrm{Im}K(\omega).
\end{align}

Finally, note that the non-zero phase shift~\eqref{eq:Phi} exists only because of operator nature of the coupling $\hat V(\tau)$ and its non-commutativity at distinct times $\tau_1 \! \neq \! \tau_2$. On the other hand, the attenuation function~\eqref{eq:chi} is given by the anticommutator, and the non-commutativity is not necessary for the result to be nonzero. In fact, within the external noise approximation to qubit--environment coupling, when one replaces operator $\hat{V}$ in $\hat H_\mathrm{int}$ with a Gaussian stochastic process $\xi(t)$ and represents the average over $E$ with the average over trajectories of $\xi$, the attenuation function remains in the same form with $\mathrm{Re}\langle \{\hat V(\tau_1),\hat V(\tau_2)\}\rangle/2\to\langle\xi(\tau_1)\xi(\tau_{2})\rangle$, while the phase simply vanishes. Thus, at least, when decoherence is well described within the Gaussian approximation~(\ref{eq:W_Gaussian}), the appearance of nonzero $\Phi(t)$ means that the environmental influence on the qubit {\it cannot} be represented by an external noise.

\section{Dephasing of NV center interacting with polarized bath} \label{sec:NV}

\subsection{The Hamiltonian of the NV center and the nuclear environment}
The Hamiltonian of spin-1 NV center interacting with a bath of spin-1/2 nuclei is given by: 
\begin{equation}
\hat  H_\mathrm{NVE}=\Omega \hat S_z+\Delta \hat S_z^2+\hat S_z \otimes\hat{V}+\hat H_E,  \label{eq:H}
\end{equation}
where $\Omega$ is the Zeeman splitting of the qubit, $\Delta$ is the zero-field splitting, $\hat H_E$ is the Hamiltonian of the environment, and $\hat{V}$ is the coupling operating in the environmental subspace. The Zeeman splitting is given by $\Omega=-\gamma_e B_0$ with $\gamma_e = \unit{28.02}{\giga\hertz\per\tesla}$ and $B_0$ is the magnetic field. The value of zero-field splitting is $\Delta=\unit{2.87}{\giga\hertz}$. The magnetic field is parallel to the quantization axis $z$ set by the zero-field splitting, that in turn, results from the physical shape of the center.

The subspace of the lowest-energy degrees of freedom  of the center is spanned by spin eigenstates $\{ |1,1\rangle,|1,0\rangle,|1,{-}1\rangle\}$, where $\hat S^2|s,m\rangle = s(s+1)|s,m\rangle$ and $\hat S_z|s,m\rangle = m|s,m\rangle$. 
Thus, we have a freedom to construct two types of effective qubits based on either $\{|1,0\rangle,|1,{\pm}1\rangle\}$ or $\{|1,1\rangle,|1,{-}1\rangle\}$ manifold. Most experiments are done employing the former, but the latter was also used in experiments on decoherence of NV centers \cite{Huang_NC11,Dolde_NP13}.

If we choose $\{|1,1\rangle,|1,{-}1\rangle\}$ manifold and assign the basis states as $|1,1\rangle \to |{\uparrow}\rangle$, $|1,{-}1\rangle \to |{\downarrow}\rangle$, then we obtain the effective qubit--environment Hamiltonian 
\begin{align}
\nonumber
\hat H_{1{-}1} &= \Omega ( |{\uparrow}\rangle\langle{\uparrow}| - |{\downarrow}\rangle\langle{\downarrow}| ) + \Delta^2 (|{\uparrow}\rangle\langle{\uparrow}|+|{\downarrow}\rangle\langle{\downarrow}|)\\
\nonumber
&\phantom{=}
	+(|{\uparrow}\rangle\langle{\uparrow}| - |{\downarrow}\rangle\langle{\downarrow}|)\otimes\hat V + \hat H_E\\
&= \Omega \hat\sigma_z + \Delta^2 \hat{\mathds{1}} + \hat\sigma_z \otimes\hat V + \hat H_E,
\end{align}
so that one gets unbiased coupling with $\lambda = \lambda_{1{-}1} = 2$ and $\eta = \eta_{1{-}1} = 0$. The second choice, $\{ |1,0\rangle, |1,{\pm}1\rangle\}$, with the assignment $|1,0\rangle\to |{\downarrow}\rangle$ and $|1,{\pm}1\rangle \to|{\uparrow}\rangle$, yields the following Hamiltonian
\begin{align}
\nonumber
\hat H_{0{\pm}1} &= (\Delta^2 \pm\Omega)|{\uparrow}\rangle\langle{\uparrow}| \pm |{\uparrow}\rangle\langle{\uparrow}|\otimes\hat V + \hat H_E\\
& = (\Delta^2 \pm \Omega)|{\uparrow}\rangle\langle{\uparrow}| \pm \frac{1}{2}(\hat{\mathds{1}}+\hat\sigma_z)\otimes\hat V + \hat H_E.
\end{align}
In this case the coupling is biased with $\eta = \eta_{0{\pm}1} = 1$ and $\lambda = \lambda_{0{\pm}1} = \pm 1$. Therefore, by switching between the manifolds we have the ability to turn the bias on and off.

\subsection{The nuclear environment}
The environment of NV center is composed of nuclear spin-$1/2$'s of ${}^{13}\mathrm{C}$ atoms. Its dynamics are modeled with the following Hamiltonian
\begin{equation}
\label{eq:H_E}
\hat H_E=\omega\sum\limits_k \hat{I}^{(k)}_z+\sum_{k<l} B_{kl} (\hat{I}^{(k)}_{+}\hat{I}^{(l)}_{-}+\hat{I}^{(k)}_{-}\hat{I}^{(l)}_{+}-4\hat{I}^{(k)}_z\hat{I}^{(l)}_z),
\end{equation}
where $\hat I^{(k)}_\alpha$ is the $\alpha$ component of spin operator of $k$th nucleus, and $\hat I_{\pm}^{(k)} = \hat I_x^{(k)} \pm i \hat I^{(k)}_y$.

The first term in \eqref{eq:H_E} describes the Zeeman splitting due to external magnetic field, with $\omega = \gamma B_0 = \unit{10.71}{\mega\hertz\per\tesla}~\times~B_0$, where $\gamma$ is the gyromagnetic ratio of ${}^{13}\mathrm{C}$ nuclei. The second term describes the secular approximation to the dipolar coupling between nuclear spin pairs, that is justified for the range of magnetic fields considered here. The pair-wise coupling strengths are given by
\begin{equation}
B_{kl}=\frac{\mu_0 \gamma^2}{4\pi |\mathbf{r}_{kl}|^3}(1-3 \cos^2\theta_{kl}),
\end{equation}
where $\mu_0$ is the magnetic permeability of the vacuum, $\mathbf{r}_{kl}$ is the vector between the positions of $k$th and $l$th nuclei, and $\theta_{kl}$ is the angle between the direction of the applied magnetic field ($z$ axis of NV center) and $\mathbf{r}_{kl}$.

\subsection{NV--environment coupling}
The coupling $\hat{V}$ describes the $z$ component of the hyperfine interaction between the electronic spin-1 of the NV center and the spin-$1/2$ of the carbon nuclei. Due to large zero-field splitting $\Delta$ (here, we consider the case when $\Delta$ dominates over Zeeman splitting $\Omega$), the influence of transverse terms (i.e.,~those projected onto $x$ and $y$ axes) on the qubit is negligible, and hence, they have been omitted. 

The explicit form of the coupling reads
\begin{equation}
\hat{V}=\sum_k\sum\limits_{\alpha=x,y,z} \aVec^{(k)}_\alpha\hat{I}^{(k)}_\alpha,
\end{equation}
with the $k$th coupling strength vector is given in the terms of the relative position between NV center and the $k$th nucleus~$\mathbf{r}^{(k)}$,
\begin{equation}
\label{eq:dip}
\mathbf{\aVec}^{(k)}=\frac{\mu_0\gamma_e\gamma}{4\pi |\mathbf{r}^{(k)}|^5}
\left(\begin{array}{c}
-3r^{(k)}_x r^{(k)}_z\\-3r^{(k)}_y r^{(k)}_z\\1-3 (r^{(k)}_z)^2\\
\end{array}\right).
\end{equation}

\subsection{Dephasing of the effective qubit}  \label{sec:evolution}
For the chosen qubit manifold $\{|1,m_1\rangle, |1,m_2\rangle\}$, the total system is initialized in a product state
\begin{align}
\hat{\rho}_{QE}(0)=|{\uparrow}_x\rangle\langle{\uparrow}_x|\otimes \hat{\rho}_E(0),
\end{align}
where $|{\uparrow}_x\rangle = (|{\uparrow}\rangle+|{\downarrow}\rangle)/\sqrt 2 = (|1,m_1\rangle + |1,m_2\rangle)/\sqrt 2$ is the superposition of the eigenstates of the effective qubit. The initial state of the environment is assumed to be prepared with a DNP procedure so that
\begin{equation}
\hat{\rho}_{E}(0)= \bigotimes_{k=1}^N \left(\frac{1}{2}\hat{\mathds{1}}+ p_k\hat{I}_z^{(k)}\right),
\end{equation}
where $p_k\in [-1,1]$ is the polarization of the $k$th spin.

We simulate the ensuing evolution of the coherence function $W_{m_1m_2}(t)$---including the spin echo realized with $\pi$ pulse applied in the midpoint of the duration---using the CCE method \cite{Yang_CCE_PRB08,Zhao_PRB12,Yang_RPP17,Kwiatkowski_PRB18}. The method allows for systematic treatment of inter-nuclear interactions by incorporating the contribution from successively larger ``clusters'' of nuclei. However, in the regime of timescales ($t\leqslant \unit{50}{\micro\second}$) and magnetic field intensities ($B_0\leqslant \unit{0.1}{\tesla}$) considered here, the influence of the intrabath dipole coupling is expected to be negligible. We have confirmed this with CCE-2 simulation, where, at most, the clusters of two nuclei are taken into account \cite{Yang_CCE_PRB08,Zhao_PRB12,Yang_RPP17,Kwiatkowski_PRB18}. Therefore, the coherence function can be simulated with a satisfactory precision even with CCE-1 approximation, where the contribution from each nuclei is treated {\it independently}. Unsurprisingly, it implies a profound consequence for the structure of the coherence function. Recall the cumulant expansion introduced in Sec.~\ref{sec:Gaussian}, Eq.~\eqref{eq:cumulant_exp}
\begin{align}
W_{m_1m_2}(t) = {\exp}\left[\sum_{n=0}^\infty \lambda_{m_1m_2}^n \kappa_n(t|f_\mathrm{echo};m_1m_2)\right].
\end{align}
(In what follows, for clarity, we will omit the dependencies in cumulants.) The fundamental property of cumulants is their additivity: the cumulant of a system composed of $N$ independent constituents decomposes into $\kappa_n = \sum_{k=1}^N\kappa_n^{(k)}$, where each $\kappa_n^{(k)}$ corresponds to the $k$th constituent. Therefore, the coherence function factorizes
\begin{align}
\nonumber
W_{m_1m_2}(t) &= e^{\sum_{n=0}^\infty \lambda_{m_1m_2}^n \kappa_n} 
	= e^{\sum_{k=1}^N \sum_{n=0}^\infty \lambda_{m_1m_2}^n \kappa_n^{(k)}}\\
\label{eq:W_CCE1}
&= \prod_{k=1}^N e^{\sum_{n=0}^\infty \lambda_{m_1m_2}^n \kappa_n^{(k)}}
	\equiv \prod_{k=1}^N W_{m_1m_2}^{(k)}(t),
\end{align}
with the single-nucleus contributions given by the following exact formula
\begin{widetext}
\begin{align}
\nonumber
W^{(k)}_{m_1m_2}(t)&= 1 - 2\lambda_{m_1m_2}^2\frac{\big(\aVec^{(k)}_x\big)^2+\big(\aVec^{(k)}_y\big)^2}{\big(\omega_{+}^{(k)}\omega_{-}^{(k)}\big)^2}\,\omega^2
	\sin^2\left(\frac{\omega_{+}^{(k)}t}{4}\right)\sin^2\left(\frac{\omega_{-}^{(k)}t}{4}\right)\\
\nonumber
&\phantom{=}
	+\frac{i}{2}\,p_k\,\lambda_{m_1m_2}^2 \frac{\big(\aVec^{(k)}_x\big)^2+\big(\aVec^{(k)}_y\big)^2}{\big(\omega_{+}^{(k)}\omega_{-}^{(k)}\big)^2}\,\omega
	\left[
		(1+\eta_{m_1m_2})\omega_{-}^{(k)}\sin\left(\frac{\omega_{-}^{(k)} t}{2}\right)\sin^2\left(\frac{\omega_{+}^{(k)}t}{4}\right)\right.\\
\label{eq:Wk}
&\phantom{=+i4p_k\,\lambda_{m_1m_2}^2 \frac{\big(A^{(k)}_x\big)^2+\big(A^{(k)}_y\big)^2}{\big(\omega_{+}^{(k)}\omega_{-}^{(k)}\big)^2}\omega}\left.
		-(1-\eta_{m_1m_2})\omega_{+}^{(k)}\sin\left(\frac{\omega_{+}^{(k)} t}{2}\right)\sin^2\left(\frac{\omega_{-}^{(k)}t}{4}\right)
	\right],
\end{align}
where
\begin{align}
\omega_{\pm}^{(k)} &= \sqrt{\tfrac{1}{4}\lambda_{m_1m_2}^2(1\pm\eta_{m_1m_2})^2\left[\big(\aVec_x^{(k)}\big)^2+\big(\aVec_y^{(k)}\big)^2\right]
	+\left(\omega\pm\tfrac{1}{2}\lambda_{m_1m_2}(1\pm\eta_{m_1m_2})\aVec_z^{(k)}\right)^2}
\end{align}
We can see that presence of non-zero polarization leads to an appearance of a term in Eq.~\eqref{eq:Wk} that is purely imaginary.
\end{widetext}

\section{Weak coupling and Gaussian approximations to NV dephasing} \label{sec:WCvsG}
The dynamics of the nuclear environment described in Sec.~\ref{sec:evolution}, is an example of a situation in which, from the point of view of the qubit, the entities constituting the environment (i.e.~nuclear spins) are approximately independent. As discussed above, the coherence of the qubit factorizes into a product of contributions from each individual constituent. It will be useful to rewrite Eq.~(\ref{eq:W_CCE1}) as
\begin{align}
\label{eq:Wprod}
W(t) = \prod_{k=1}^N W^{(k)}(t) \equiv \prod_{k=1}^N [1 - \delta W^{(k)}(t)] \,\, ,
\end{align}
where we have defined $ \delta W^{(k)}(t)$.
For clarity, we will omit the effective qubit manifold subscript in this section.

Let us now consider the case in which every constituent of the environment only weakly perturbs the qubit $|W^{(k)}(t)|\approx 1$, or equivalently $|\delta W^{(k)}(t) | \ll 1$ for every $k$; we have then
\begin{align}
\label{eq:Weak}
W(t) = e^{ -\sum_k \delta W^{(k)}(t)}[1 + O(\delta W^2)].
\end{align}
We will refer to this as a {\it weak-coupling approximation} result. 

For $W^{(k)}(t)$ given by Eq.~(\ref{eq:Wk}), the weak coupling condition is given by
\begin{align}
\epsilon_k \equiv \frac{\sqrt{\big(\aVec_{x}^{(k)}\big)^2+\big(\aVec_y^{(k)}\big)^2}}{\omega}\ll 1,
\end{align}
as it guarantees that the amplitudes of all the oscillatory terms in Eq.~(\ref{eq:Wk}) are small. When this is fulfilled we have
\begin{align}
\nonumber
W_{\mathrm{weak}}(t) & \approx e^{\sum_k[W^{(k)}(t)-1]} = e^{-\sum_k\delta W^{(k)}(t)}\\
\nonumber
&= e^{-\sum_k\mathrm{Re}\big\{\delta W^{(k)}(t)\big\}}e^{-i\sum_k \mathrm{Im}\big\{\delta W^{(k)}(t)\big\}}\\
\label{eq:W_weak}
&\equiv \prod_{k=1}^N W^{(k)}_\mathrm{weak}(t),
\end{align}
where the single-nucleus contribution is given by
\begin{align}
\label{eq:Wk_weak}
W^{(k)}_\mathrm{weak}(t) = e^{-\lambda^2 \mathcal{R}^{(k)}(t)} e^{-i p_k\lambda^2\mathcal{I}^{(k)}(t)}.
\end{align}
A specific feature of the weak-coupling approximation is that the polarization-induced phase $\mathcal{I}^{(k)}(t)$ is determined {\it only} by the imaginary part of $\sum_k\delta W^{(k)}$, and it does not affect the coherence when the polarization $p_k$ is zero. At the same time, $\mathcal{R}^{(k)}$ is determined by the real part and it is {\it independent} of $p_k$. Of course, this ceases to be the case when the approximation breaks down.

On the other hand, if instead of weak-coupling, one applies the Gaussian approximation to each constituent, then
\begin{align}
\label{eq:Wk_Gauss}
W^{(k)}_\mathrm{Gauss}(t) = e^{-\lambda^2\chi^{(k)}(t)}e^{-i\eta\lambda^2 \Phi^{(k)}(t)},
\end{align}
with the attenuation function and bias-induced phase shift given by Eqs.~\eqref{eq:chi} and \eqref{eq:Phi}, which, in the case of nuclear environment consisting of noninteracting spins, read
\begin{align}
\chi^{(k)}(t) &= 2\epsilon_k^2\sin^4\left(\frac{\omega t}{4}\right),\\
\label{eq:Gaussian_phase}
\Phi^{(k)}(t) &= 2 p_k\,\eta\,\epsilon_k^2\cos\left(\frac{\omega t}{4}\right)\sin^3\left(\frac{\omega t}{4}\right).
\end{align}
Clearly, the results of weak-coupling \eqref{eq:Wk_weak} and Gaussian \eqref{eq:Wk_Gauss} approximations do not coincide, e.g., the Gaussian phase shift is proportional to the bias $\eta$, while weak-coupling predicts a complicated non-linear dependence. The main differentiating factor between the two approximations are the frequencies within the corresponding oscillating terms
\begin{align}
\nonumber
\omega^{(k)}_\pm \! &= \omega\sqrt{1\pm(1\pm\eta)\lambda\frac{\aVec^{(k)}_z}{\omega}+\frac{(1\pm\eta)^2\lambda^2}{4}\frac{(\mathbf{\aVec}^{(k)})^2}{\omega^2}}\\
\nonumber
&=\omega \pm \frac{(1\pm\eta)\lambda\, \aVec_z^{(k)} }{2} + O\Big(\!(1\pm \eta)^2\lambda^2\frac{ (\aVec_\alpha^{(k)})^2}{\omega}\!\Big)\\
&\equiv \omega + \delta\omega_\pm^{(k)}.
\end{align}
Consequently, the Gaussian and weak-coupling coherence functions will eventually {\it desynchronize}---they will pass through their local extrema at different times, and the delay will only increase over the number of cycles.

When compared on a level of the contribution from single nucleus, the two approximations will not converge naturally. Indeed, the Gaussian approximation is typically justified with the central limit theorem, that has two essential requirements for it to work: (i) the independent constituents have to be weakly-coupled (which we already assume), (ii) the number of constituents have to be large. The second point implies that we should investigate the impact of $N$---the number of nuclei coupled to the effective qubit. To this end, let us examine the coherence function as a whole
\begin{align}
|W_\mathrm{Gauss}(t)|= \Big|\prod_{k=1}^N W^{(k)}_\mathrm{Gauss}(t)\Big| \leqslant e^{-2\lambda^2\epsilon_0^2 N\sin^4\left(\frac{\omega t}{4}\right)}
\end{align}
where $\epsilon_0 = \min_{k}\epsilon_k$. As we can see, even though each single-nucleus contribution $W^{(k)}(t) \! \approx \! 1$, as we already assumed that $\epsilon_0\ll 1$, the cumulative effect, indicated by the amplifying factor $N\gg 1$, is stronger by orders of magnitude. The same is true for collective contribution from $W^{(k)}_\mathrm{weak}(t)$. Hence, if we wish to keep the magnitude of the coherence on the level where it can still be observed, then we have to restrict the duration of the evolution $t$ so that the time-dependent part of the attenuation function can temper the amplification from the number of nuclei $N$. To quantify this, let us define the time-scale $t_c$ on which the magnitude of the coherence function is still appreciable
\begin{align}
N\sin^4\left(\frac{\omega t_c}{4}\right) = 1 \quad\Rightarrow\quad t_c \leqslant \frac{4 N^{-\frac{1}{4}}}{\omega} \ll \frac{1}{\omega}.
\end{align}
Then, on this time-scale, the frequency differences are negligible
\begin{align}
\frac{\delta\omega_{\pm}^{(k)} t_c}{4} &= \pm (1\pm\eta)\lambda N^{-\frac{1}{4}}\frac{\aVec^{(k)}_z}{\omega} + O\left(\frac{(\aVec_\alpha^{(k)})^2}{N^{\frac{1}{4}}\omega^2}\right)
	\ll 1,
\end{align}
and both $\mathcal{R}^{(k)}(t) = \chi^{(k)}(t)$ and $p_k\mathcal{I}^{(k)}(t)=\Phi^{(k)}(t)$ in the lowest-order of $\aVec^{(k)}_\alpha/\omega$.

Beyond the time-scale $t_c$, the coherence function sharply decays below the measurability threshold, where any differences between approximation schemes are irrelevant. However, this changes when the duration $t$ is allowed to reach a multiple of the period of the oscillatory terms and the coherence function revives (close) to its initial state. For Gaussian approximation, these revival times are given by
\begin{align}
\frac{\omega t_n}{4} = n \pi\quad\Rightarrow\quad t_n = n \frac{4\pi}{\omega},
\end{align}
where $n$ is a natural number. When compared with the weak-coupling we get
\begin{align}
\frac{(\omega+\delta\omega_\pm^{(k)})t_n}{4} = n\pi \pm \pi(1\pm\eta)\lambda\, n\frac{\aVec^{(k)}_z}{\omega} +O\left(\!n\frac{(\aVec^{(k)}_\alpha)^2}{\omega^2}\!\right),
\end{align}
which means that the desynchonization between revivals of Gaussian and weak-coupling results would become noticeable after $n\sim (\aVec^{(k)}_z/\omega)^{-1}$ cycles. Moreover, since the frequencies $\delta\omega_\pm^{(k)}$ depend on the magnitude of couplings $\aVec_\alpha^{(k)}$, the revival timings are desynchonizing at sightly different rates. As a result, the shape of the revival peak (its width, height, kurtosis, etc.) will be distorted in comparison to ideally periodic revivals present within Gaussian approximation. Nevertheless, each revival event---i.e., when the coherence raised from below the measurability threshold up to its local maximum, and then back below the threshold---takes place over time-scale~$t_c$. Hence, while they remain in sync, the Gaussian and weak-coupling approximations predict the same course of the evolution when the number of nuclei is large.

\section{Results for echo signal in presence of polarized nuclear environment} \label{sec:results}
We consider a diamond sample with natural concentration of spinful nuclei, i.e., the ratio of $^{13}\mathrm{C}$ isotope to $^{12}\mathrm{C}$ is on the level of~$1.1$\%. The magnetic field is set to $B_0\! =\! \unit{15.6}{\milli\tesla}$---the standard intensity value used in experiments with the nuclear environment, when one wants to focus on single-nucleus dynamics effects on dephasing  \cite{Staudacher_Science13,Haberle_NN15,DeVience_NN15,Hernandez_PRB18}. For such a field intensity, the spin echo signal exhibits prominent oscillations, or revival cycles \cite{Childress_Science06,Zhao_PRB12}, that occur with a period of $\sim \! \unit{10}{\micro\second}$; the extent of evolution duration considered here encompasses up to three such periods. As we have discussed in Sec.~\ref{sec:evolution}, we have verified with CCE-2 simulations that, on this time scale, the inter-nuclear interactions can be omitted, and each nucleus can be treated as an independent contributor. Due to the strong dependence of the NV--nucleus coupling on the distance between them, on this timescale we can neglect the contributions from nuclei located further than $\unit{10}{\nano\meter}$ from the NV center, which means that, on average, there are $N\sim 600$ contributing nuclei.

\begin{figure*}[ht]
\begin{center}
	\begin{tabular}{ccccc}
		&(I)&(II)&(III)&(IV)\\
		(a)&&&&\\
	&	\includegraphics[width=4cm]{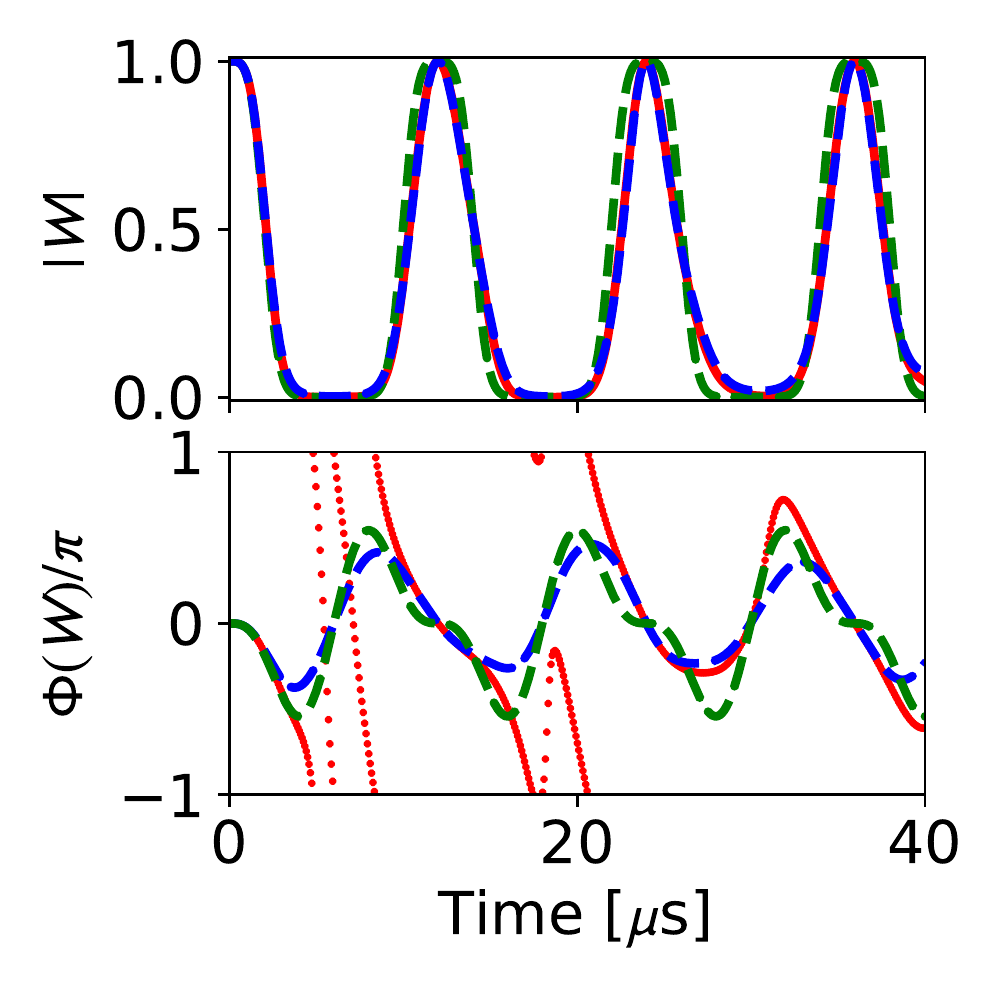}&
		\includegraphics[width=4cm]{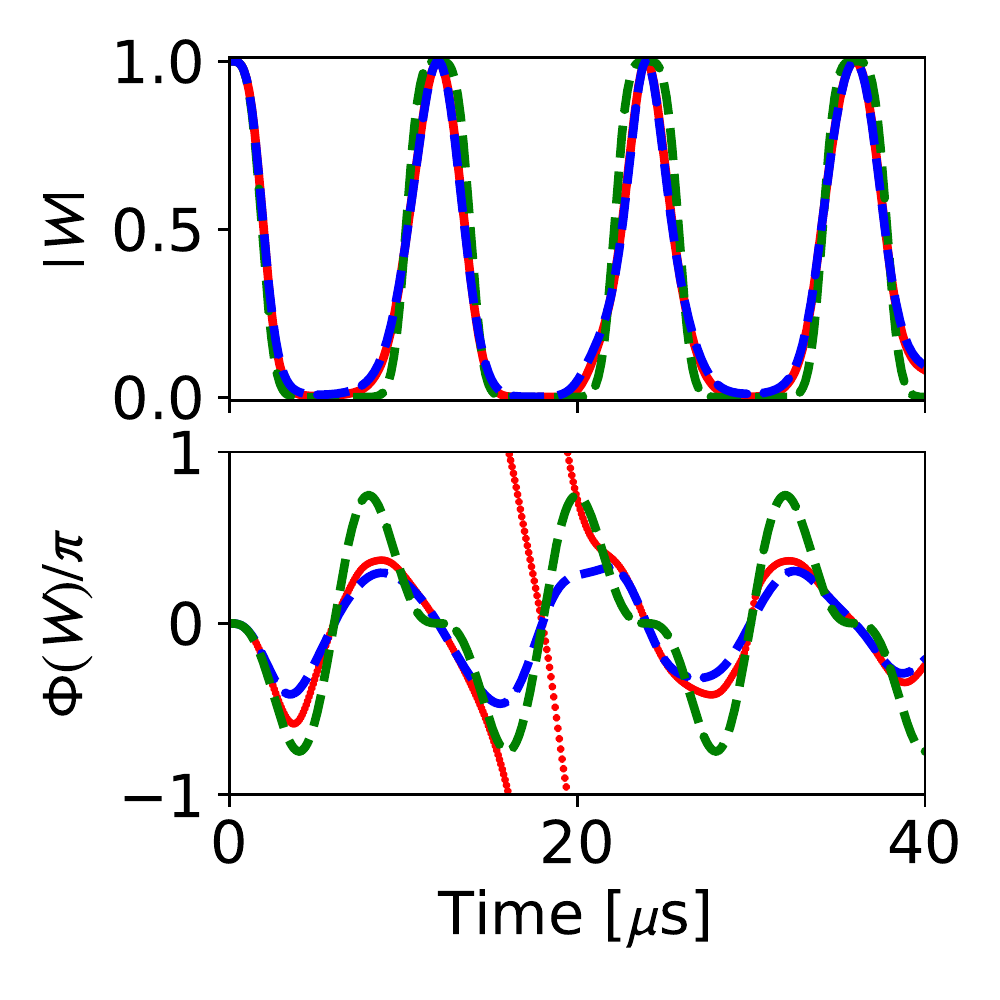}&
		\includegraphics[width=4cm]{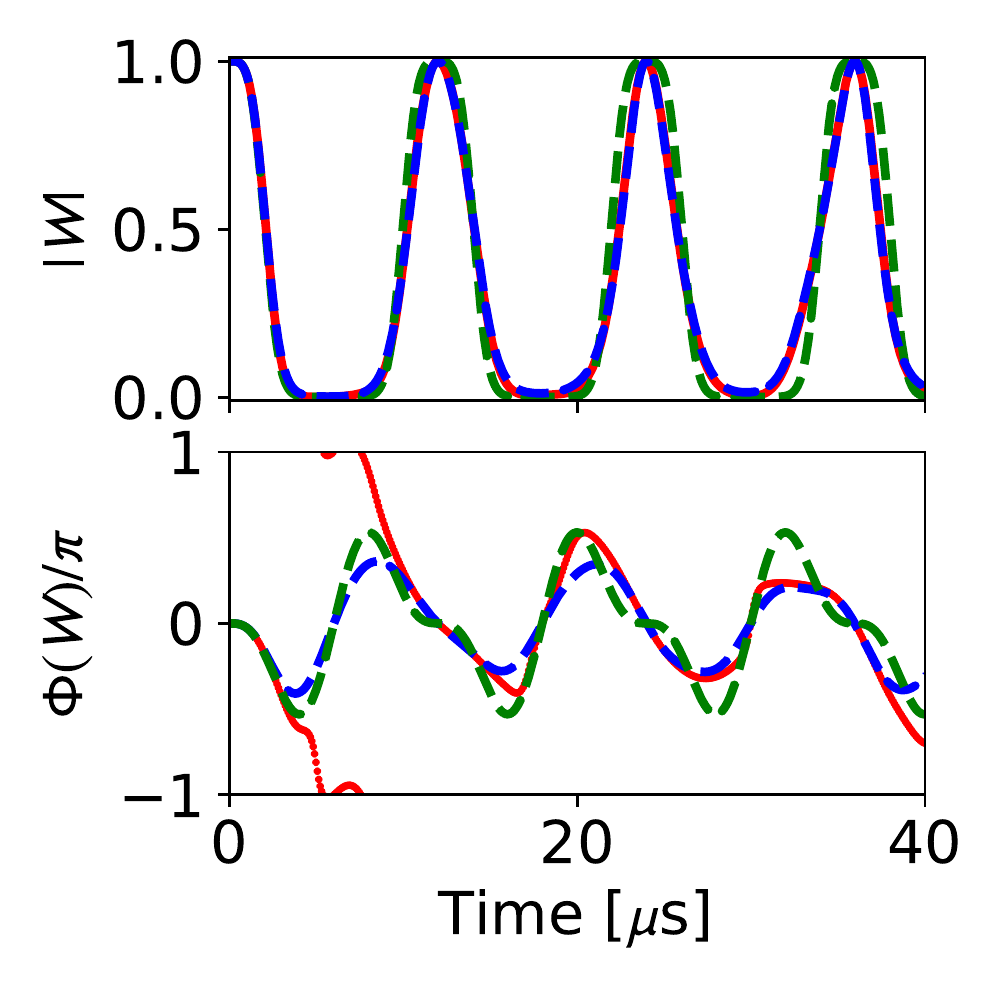}&
		\includegraphics[width=4cm]{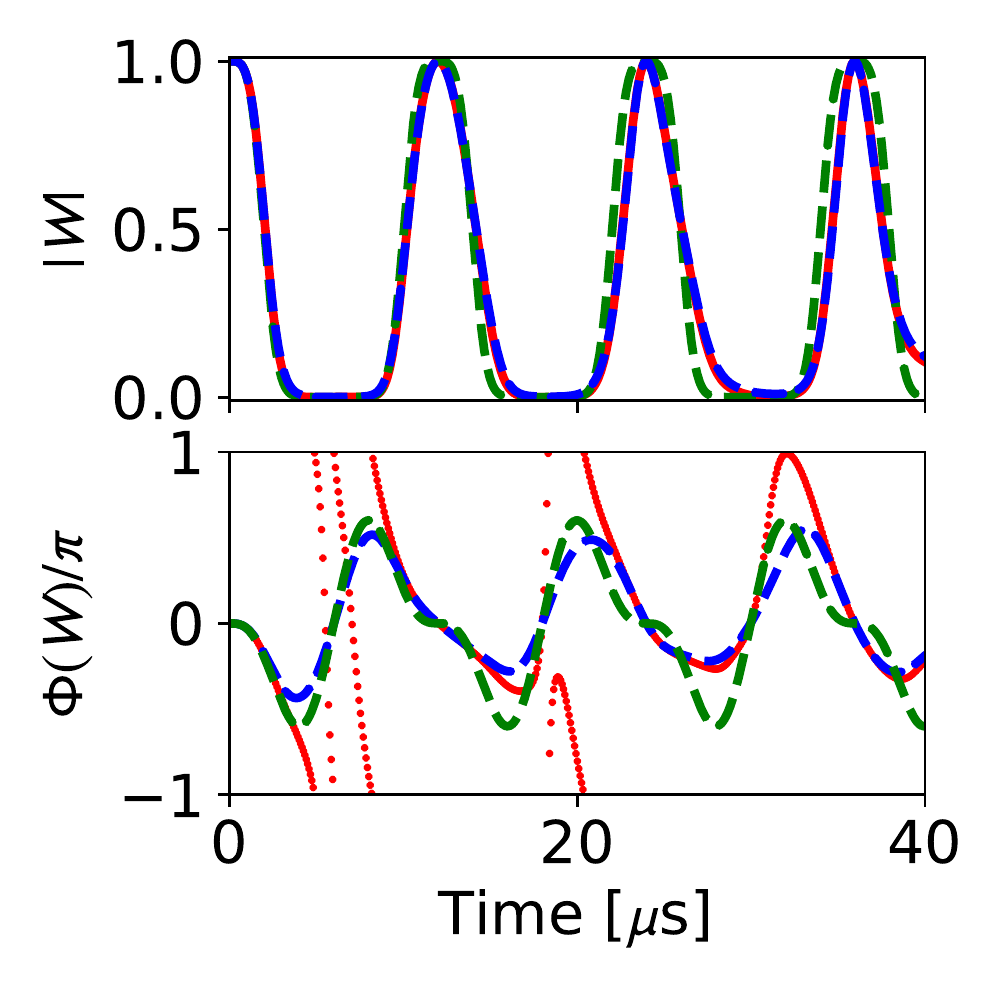}\\
		
		(b)&&&&\\
	&	\includegraphics[width=4cm]{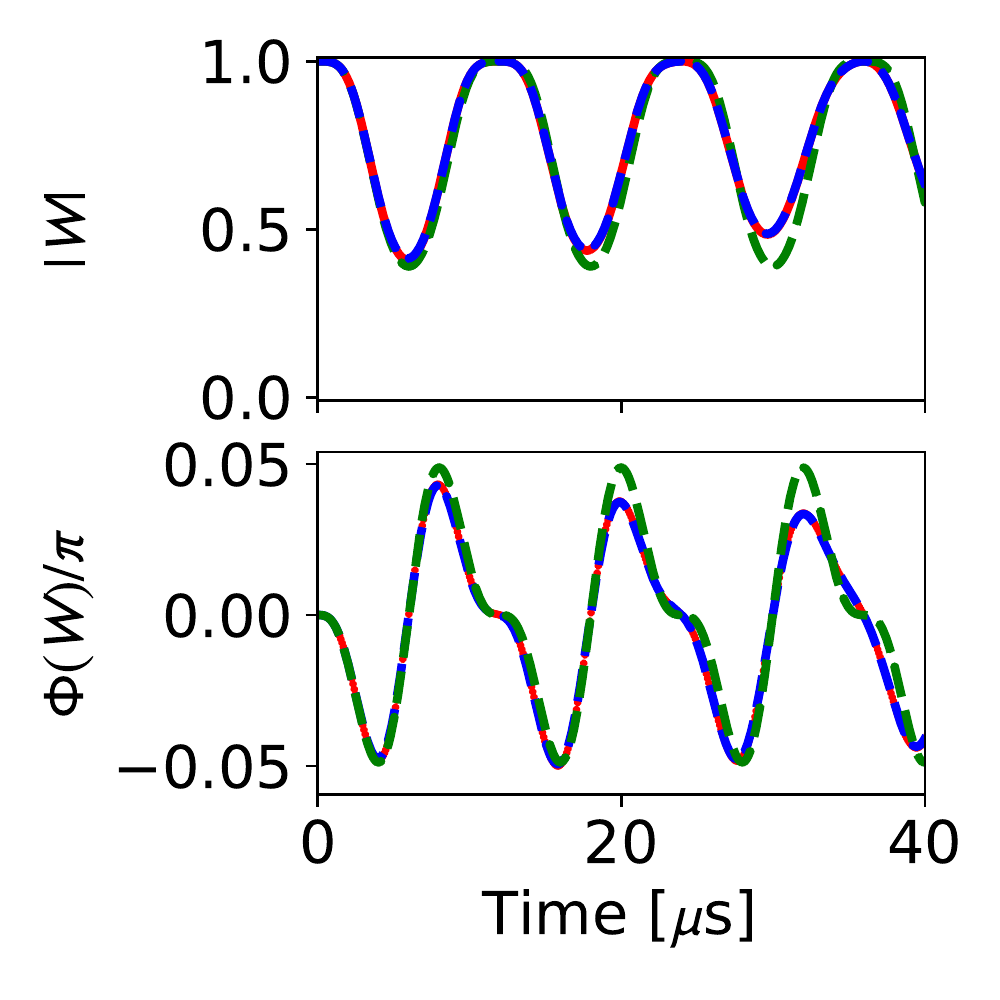}&
		\includegraphics[width=4cm]{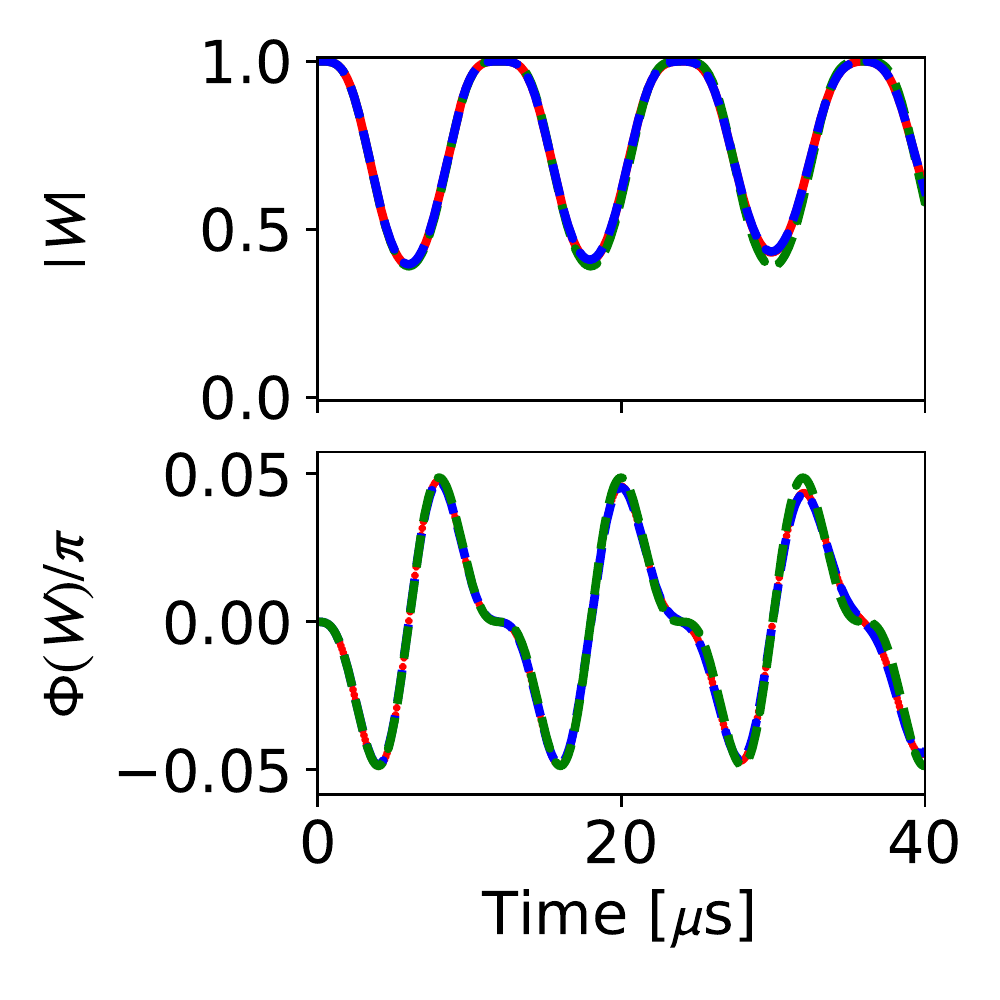}&
		\includegraphics[width=4cm]{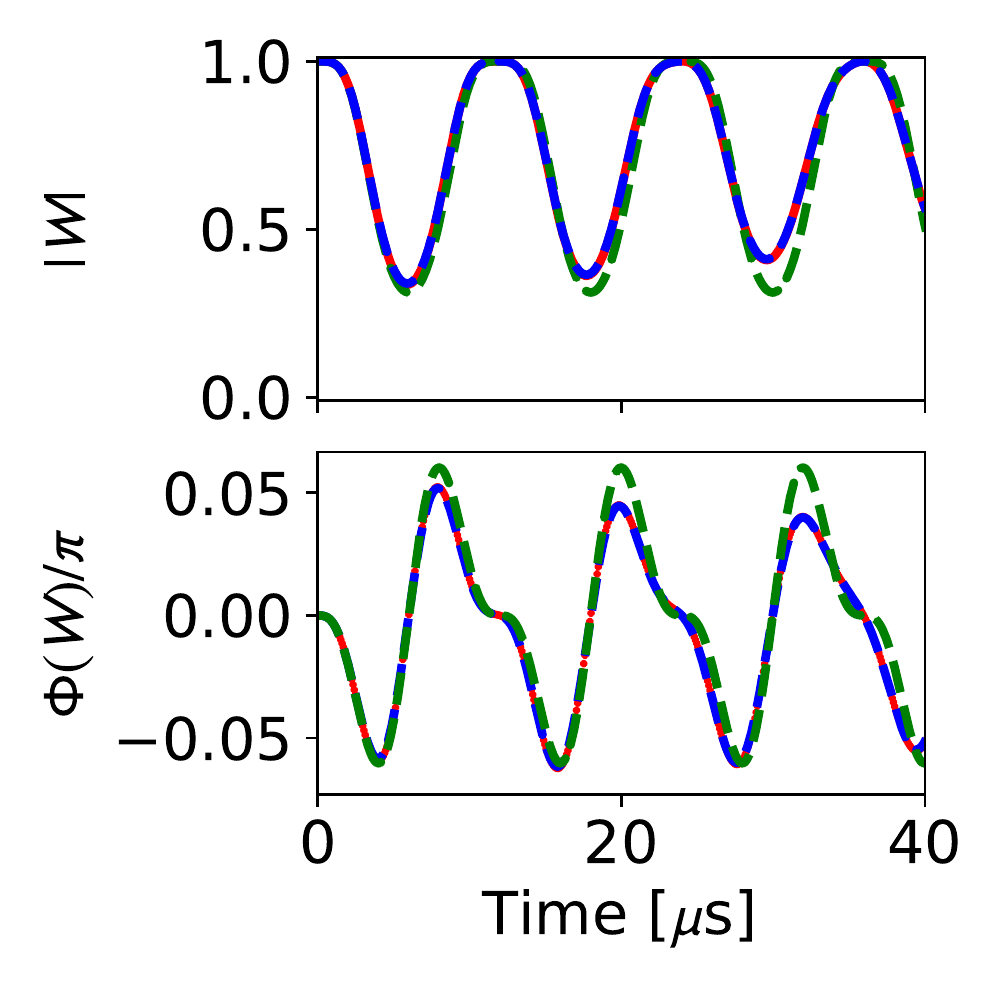}&
		\includegraphics[width=4cm]{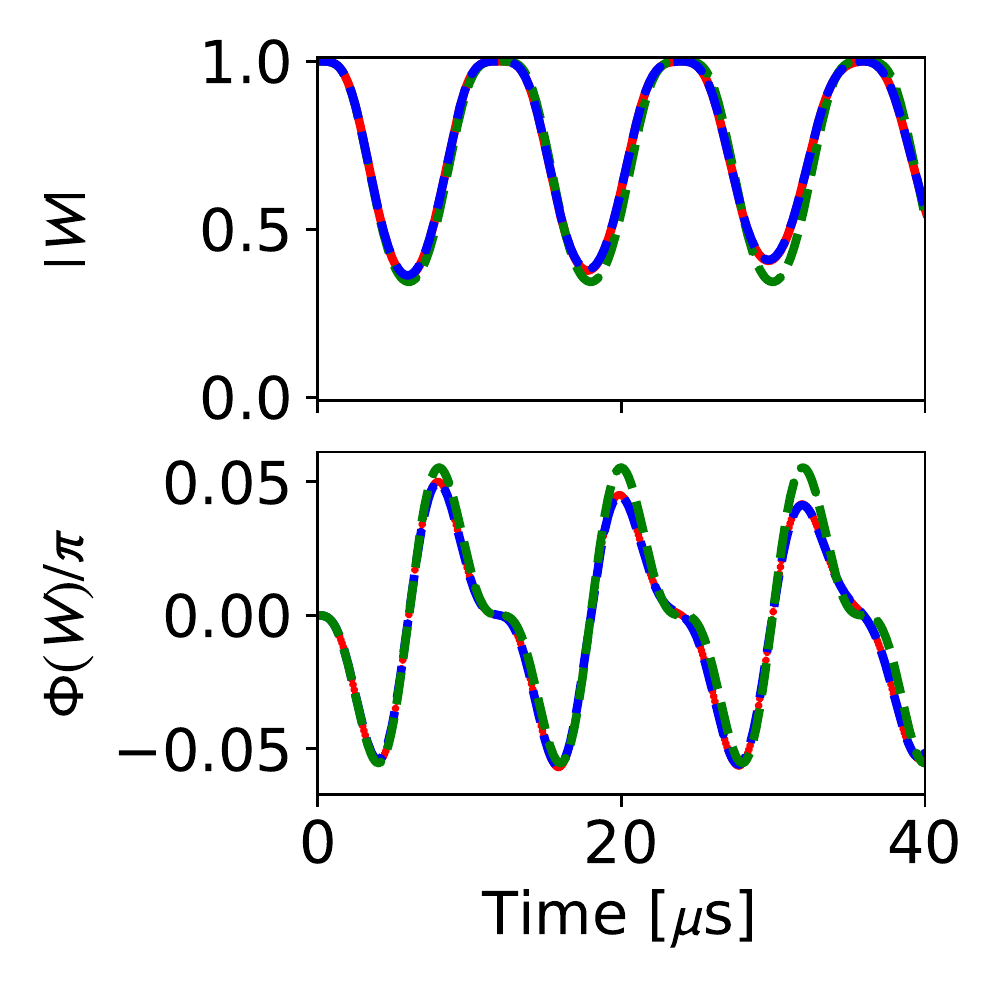}
	\end{tabular}
	\end{center}
	\caption{The modulus $|W|$ and the phase $\Phi$ of the spin echo coherence function of the effective qubit based on $\{|1,0\rangle,|1,1\rangle\}$ manifold of the NV center. Each plot compares the results of the exact CCE-1 given by Eqs.~\eqref{eq:W_CCE1} and \eqref{eq:Wk} (red dotted line), the weak-coupling \eqref{eq:Wk_weak} (blue dashed line), and the Gaussian \eqref{eq:Wk_Gauss} (green dashed line) approximations. The columns correspond to the spatial realizations I-IV of the nuclear environment, and in each case all environmental spins are polarized $p_k=0.5$ for each $k$. In the row (a) there are no nuclei with $\unit{0.5}{\nano\meter}$ distance from the qubit (the original realization), and in the row (b) the realizations have been modified---the moderately coupled nuclei within $\unit{1}{\nano\meter}$ distance form the qubit have been removed.}\label{fig:comparison_cuts}
\end{figure*}

The positions of nuclear spins within the crystal lattice are chosen at random from uniform probability distributions. In our simulations we utilize four spatial realizations (enumerated with Roman numerals I-IV) drawn in such a way. The key feature of these realization is that there are no nuclei ``nearby'' the NV center. The most interesting, and simultaneously, the most challenging aspect of nuclear environment is the difficulty in defining what ``nearby'' actually means. The culprit is the dominant NV--nucleus dipole interaction, that is not only anisotropic, but also has a power-law dependence on the distance [see Eq.~\eqref{eq:dip}]. Because of that, it is impossible to define a sharp border between ``nearby'' and ``far away'' distance scales. Therefore, we had to resort to the trial-and-error method, where by discarding those realizations for which the removal of single strongly coupled nearest neighbor of the center caused a substantial difference in the decoherence process, we have found the I-IV realizations and determined that ``nearby'' is, no less, than the distance around $\unit{0.5}{\nano\meter}$. For natural concentration of $^{13}\mathrm{C}$ spins, about $50\%$ of random realizations count as being free of strongly coupled nearest neighbors. In the remaining cases, the NV center together with one (or more) nearest neighbors should rather be treated as a multi-qubit register \cite{Dutt_Science07,Jiang_Science09,Robledo_Nature11,Taminiau_NN14,Waldherr_Nature14,Bradley_PRX19}, rather than a typical example of qubit undergoing decoherence process caused be a mesoscopic nuclear bath. Since, we are interested in the latter, we only consider realizations without strongly coupled nuclei, or, equivalently, the 
realizations with only {\it weakly} coupled nuclei.

\subsection{Tests of the accuracy of Gaussian approximation}\label{sec:gauss_test}
In Fig.~\ref{fig:comparison_cuts} we compare the exact CCE-1 echo signal $W_{01}(t)=\prod_k W^{(k)}_{01}(t)$ given by~\eqref{eq:Wk}, with its weak-coupling~\eqref{eq:Wk_weak} and Gaussian~\eqref{eq:Wk_Gauss} approximations. The environment realizations are initialized in a polarized state with $p_k=0.5$ for each $k$. Note that, for the effective qubit based on $\{|1,0\rangle,|1,1\rangle\}$ manifold, the coupling is biased $\eta_{01}=1$ and $\lambda_{01}=1$. 

As we have discussed previously, the spatial realizations I-IV consist of only weakly coupled nuclei, and hence, the weak-coupling approximation predicts the behavior of the modulus with high accuracy. The prediction of the phase shift is comparably accurate, except for the regions where the coherence drops to zero. However, those exceptions are irrelevant, because in those cases, it is impossible to accurately estimate the phase from any measurement in the first place.

As expected based on considerations from Sec~\ref{sec:WCvsG}, the Gaussian and the weak-coupling approximations match up closely while the coherence undergoes the initial decay, and start to differ over the subsequent revival cycles. We have found that the duration scale over which the Gaussian approximation remains accurate is substantially extended when the environment is also purged of the ``moderately'' coupled nuclei [row (b) in the figure]. In practice, we have modified the spatial realizations of the environment by removing all nuclei within $\unit{1}{\nano\meter}$ distance from the qubit (by doing so, we have effectively moved the border between ``nearby'' and ``far'' away from the NV center).

\subsection{Biased vs. unbiased couplings}\label{sec:bias_vs_no_bias}
\begin{figure}[tbh]
	\includegraphics[width=0.85\linewidth]{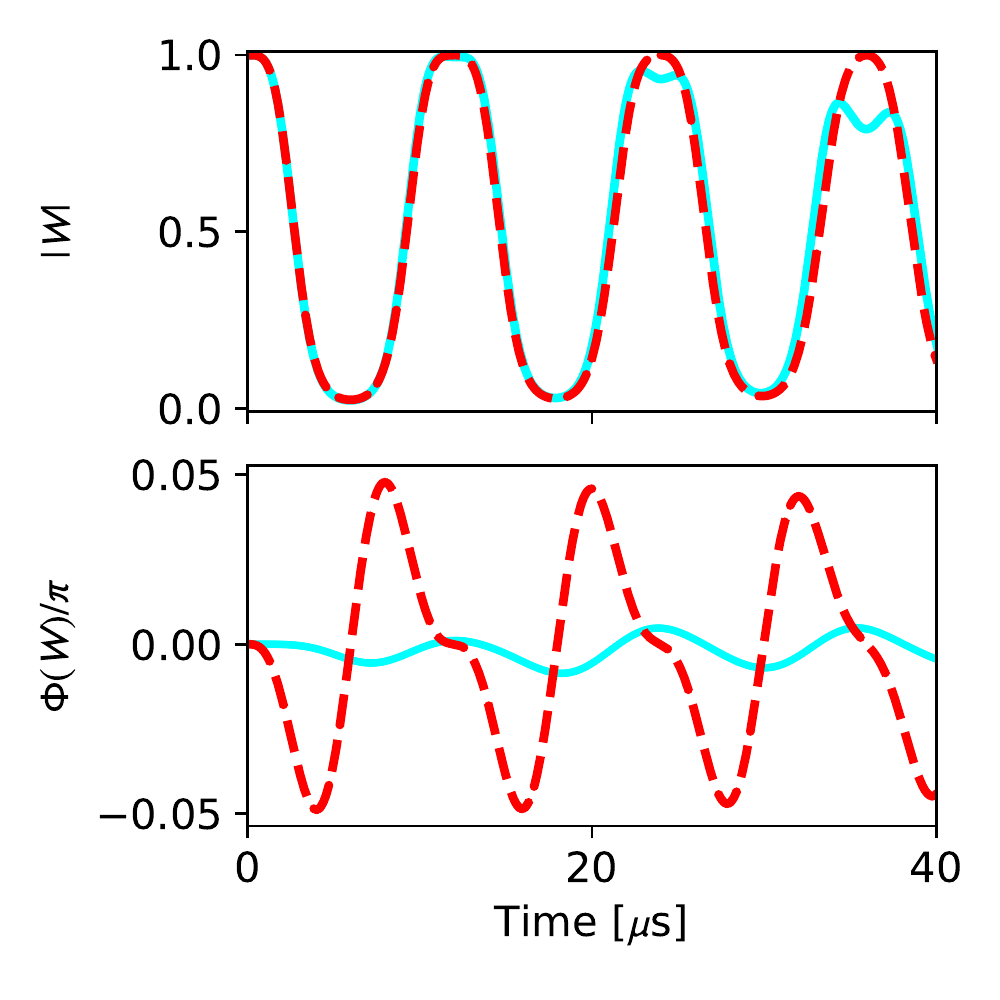}
	\caption{The comparison between spin echo coherence functions of qubits based on $\{|1,0\rangle,|1,1\rangle\}$ (biased coupling) and $\{|1,1\rangle,|1,{-1}\rangle\}$ (unbiased coupling). The environment is set in II. spatial realization with the moderately coupled nuclei removed (no nuclei with $\unit{1.0}{\nano\meter}$ radius around the NV center). Cyan solid line depicts the modulus and the phase of $W_{1{-1}}(t)$, while the red dashed line stands for modulus and phase of $W_{01}^4(t)$ (the coherence function is corrected for the difference between coupling parameters $\lambda_{1{-1}}=2\lambda_{01}$).}
\label{fig:Wpm}
\end{figure}

In Fig.~\ref{fig:Wpm} we compare the echo signals from qubits based on $\{|1,0\rangle,|1,1\rangle\}$ and $\{|1,1\rangle,|1,{-}1\rangle\}$ manifolds for fully polarized spatial realization II with the moderately coupled nuclei removed. Since $\lambda_{1{-}1} = 2\lambda_{01}$, assuming that the Gaussian approximation holds up, the modulus $|W_{1{-}1}(t)|\approx |W_{01}(t)|^4$ (note that this expectation was analyzed first in \cite{Zhao_PRL11}). Simultaneously, within the same approximation, there should be no phase shift in $W_{1{-}1}(t)$ because $\Phi_{1{-1}}(t)\propto \eta_{1{-1}}=0$. These predictions (and simultaneously, the validity of Gaussian approximation) are largely confirmed by the results shown in the figures. The moduli (corrected for the difference in $\lambda$) go in lock-step until the second revival of $|W_{1{-1}}(t)|$ where the Gaussian approximation loses its accuracy due to the desynchronization effect described in Sec.~\ref{sec:WCvsG}. As for $W_{01}(t)$, based on our previous conclusions (see Sec.~\ref{sec:gauss_test}), in this case, the Gaussian approximation holds firmly throughout. Therefore, the phase depicted in the figure can be identified with the bias-induced $\Phi_{01}(t)$. On the other hand, the phase shift in $W_{1{-1}}(t)$ indicates the deviation from Gaussian approximation, that is initially small and then increases over the course of subsequent revival cycles. Thus, we have obtained a convenient single-qubit non-Gaussianity witness: if one observes a non-zero phase shift in the echo signal from qubit based on $\{|1,1\rangle,|1,{-1}\rangle\}$ manifold, then the signal for the environment is not characterized by Gaussian statistic, and the larger the phase amplitude, the larger the deviation from Gaussianity.

\begin{figure}[tbh]
	\includegraphics[width=0.9\linewidth]{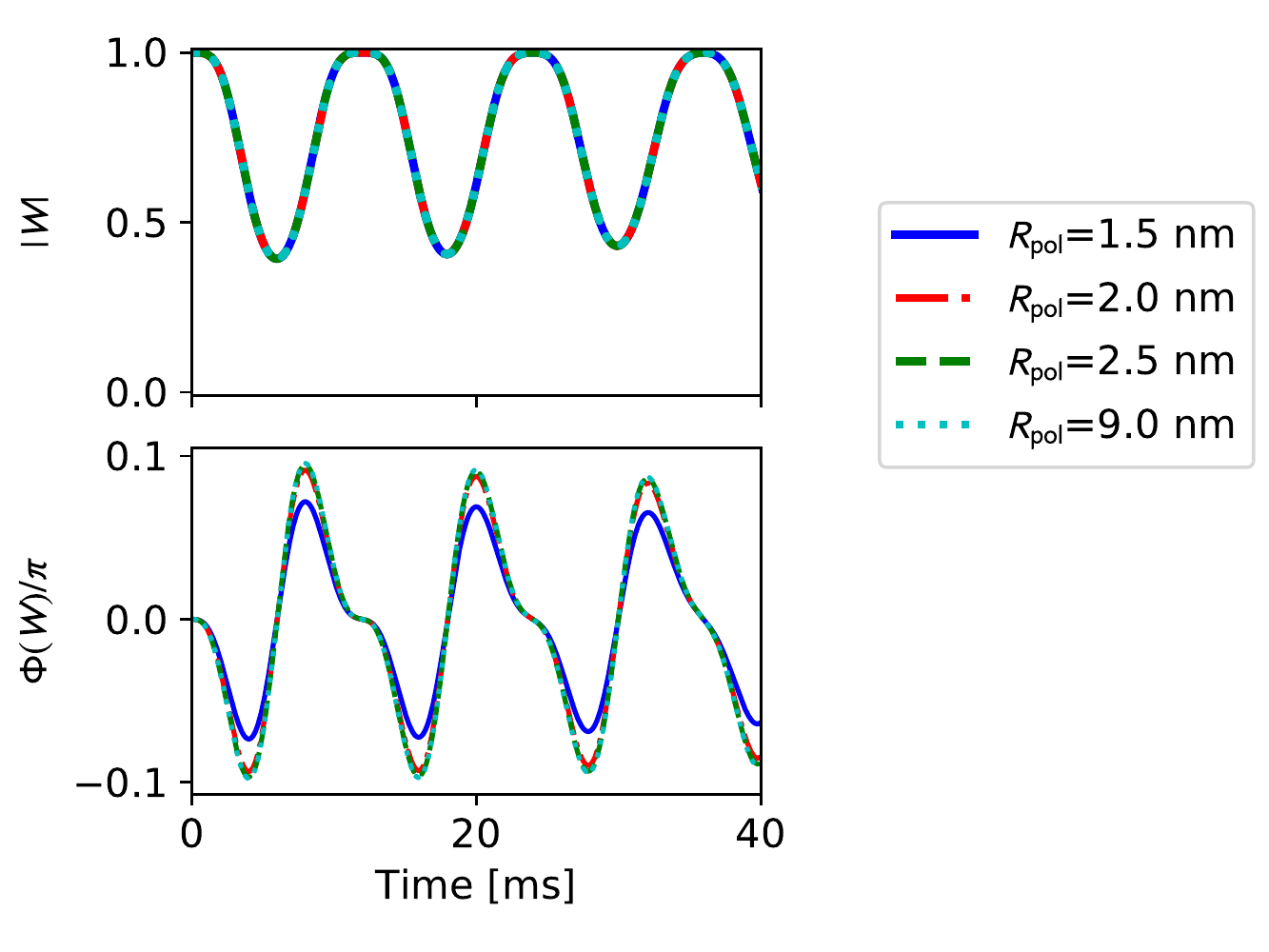}
	\caption{The effects of limited DNP range on the spin echo coherence function of qubit based on $\{|1,0\rangle,|1,{-1}\rangle\}$ manifold of the NV center (the biased coupling). The environment is set in the II. spatial realization with the moderately coupled nuclei removed (no nuclei with $\unit{1.0}{\nano\meter}$ radius around the center). Plots compare the modulus and the phase depending on the distance $R_\mathrm{pol}$ from the qubit, within which the nuclei are fully polarized: $R_{\mathrm{pol}}=\unit{1.5}{\nano\meter}$ (blue), $\unit{2.0}{\nano\meter}$ (red), $\unit{2.5}{\nano\meter}$ (green), and $\unit{9.0}{\nano\meter}$ that encompasses the whole bath (cyan).}
		\label{fig:Rpol}
\end{figure}

\subsection{The influence of the DNP range}
In practice, the DNP procedure allows to polarize the initial state of the nuclei only within certain distance $R_\mathrm{pol}$ from the NV center. Here, we investigate how this limit on the polarization range influences the echo signal $W_{01}(t)$. Specifically, we consider II. spatial realization with moderately coupled nuclei removed (i.e., no nuclear spin within \unit{1}{\nano\meter} from the NV center) where the nuclei within $R_{\mathrm{pol}}$ from the center are fully polarized , i.e.~their $p_k=1$, while those outside this range are left unpolarized with $p_k=0$. Figure~\ref{fig:Rpol} shows the modulus and the phase of $W_{01}(t)$ for four progressively larger polarization ranges $R_\mathrm{pol}$, up to $R_\mathrm{pol}=\unit{9}{\nano\meter}$ which encompasses the whole bath.

The results given in Sec.~\ref{sec:gauss_test} tell us that the Gaussian approximation holds up in this case, and so, the polarization of nuclei is expected to affect only the phase of $W_{01}(t)$---this is confirmed in the plot of $|W_{01}(t)|$ which is essentially independent of the value of $R_\mathrm{pol}$. On the other hand, the amplitude of the phase increases as the distance $R_\mathrm{pol}$ is extended and more nuclei are allowed to contribute their polarization. The trend saturates when $R_\mathrm{pol}$ extends past $\unit{2}{\nano\meter}$ and polarizing more distant nuclei does not lead to a visible modification of the phase. We see, thus, that the polarization of nuclei closer with $\approx \! \unit{2}{\nano\meter}$ radius around the NV center fully determines the amplitude of the phase of the echo signal.

\subsection{Measurability of the bias-induced phase shift}
\begin{figure*}[tbh]
	\begin{center}
		\begin{tabular}{ll}
			(a)& (b) \\
			\includegraphics[width=0.45\linewidth]{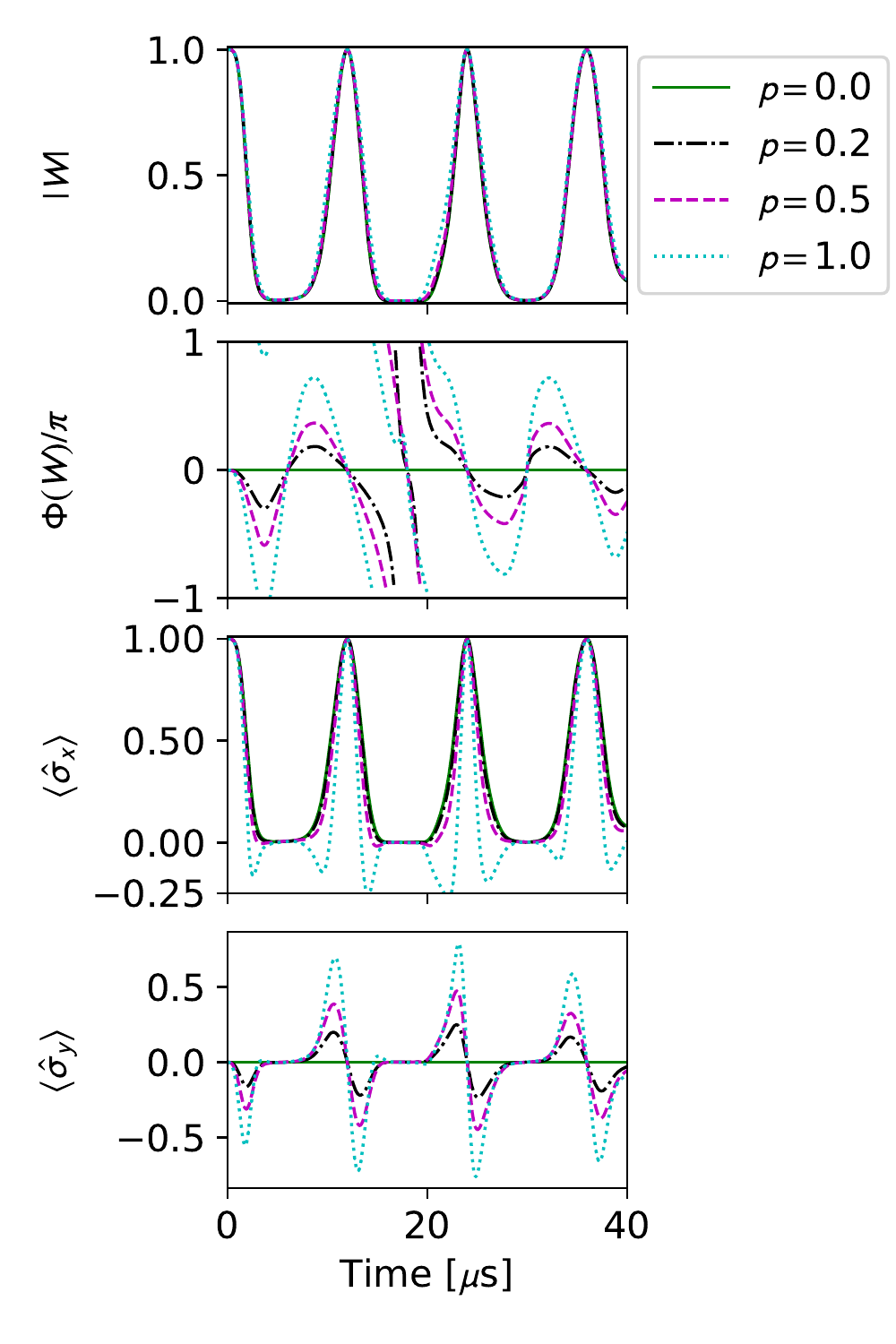} &
			\includegraphics[width=0.45\linewidth]{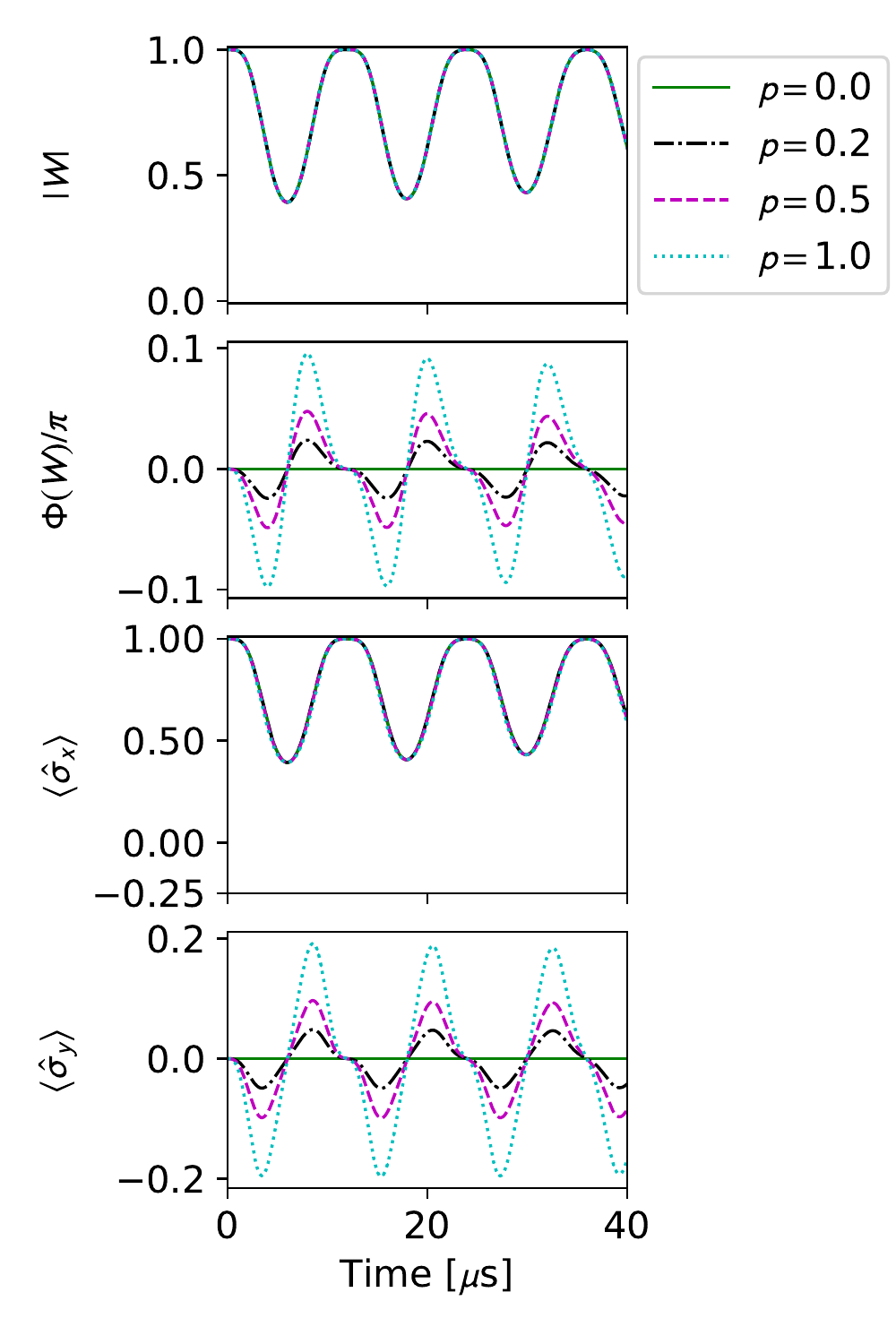} \\
		\end{tabular}
	\end{center}
	\caption{The expectation values of transverse spin components $\langle \hat\sigma_{x/y}(t)\rangle$ of qubit based on $\{ |1,0\rangle,|1,1\rangle\}$ manifold (biased coupling) for the spin echo dynamics. (The modulus and the phase of qubit's coherence function are shown for reference.) The environment is set in II. spatial realization, with the moderately coupled nuclei retained [column (a)] or removed [column (b)]. The plots compare the results obtained for different degree of polarization $p$ of the bath ($p_k=p$ for all $k$): $p=0.0$ (green), $0.25$ (red), $0.5$ (green), and $1.0$ (cyan).}\label{fig:polarization}
\end{figure*}
In order to gain access to the spin echo signal in practice, one typically measures the transverse components of qubit's spin $\langle S_{x}(t)\rangle \propto\langle\hat\sigma_{x}(t)\rangle\propto \mathrm{Re}\{W_{m_1m_2}(t)\}$ or $\langle S_y(t)\rangle \propto \langle \hat\sigma_y(t)\rangle\propto\mathrm{Im}\{W_{m_1m_2}(t)\}$. For qubit initialized in $|{\uparrow}_x\rangle = (|{\uparrow}\rangle + |{\downarrow}\rangle)/\sqrt{2}$ state, and the environment in a completely mixed state (zero polarization), the measured spin components have a very simple form $\langle\hat\sigma_x(t)\rangle = W_{m_1m_2}(t) = |W_{m_1m_2}(t)|$ and $\langle\hat\sigma_y(t)\rangle = 0$. However, when the environment is polarized, $W_{m_1m_2}(t)$ is no longer purely real, $\langle\hat\sigma_y(t)\rangle$ is non-zero, and the course of $\langle \hat\sigma_x(t)\rangle$ is modified. 

Here, we investigate to what extent this effect can be measured for the qubit based on $\{ |1,0\rangle,|1,1\rangle\}$ manifold (it is much weaker for $\{|1,1\rangle,|1,{-1}\rangle\}$ manifold because of the unbiased coupling). In Fig.~\ref{fig:polarization} we showcase the results obtained for II. spatial realization with the moderately coupled nuclei retained [column (a)] and removed [column (b)], and for a spread of bath polarization values (all nuclei equally polarized). As we can see, when the contribution from moderately coupled nuclei is retained (and thus, non-Gaussian effects become important), as in column (a), there is a strong signature of large polarization in $\langle \hat{\sigma}_x(t)\rangle$: it acquires negative values on both sides of the revival peaks. When the moderately coupled nuclei are removed, and Gaussian approximation is in effect [column (b)], $\langle \hat{\sigma}_x(t)\rangle\approx |W_{01}(t)|$, as if the bath was not polarized at all. On the other hand, the effects of polarization on $\langle \hat{\sigma}_y(t)\rangle$ are clearly visible within and beyond the regime of Gaussian approximation.

\section{Conclusion} \label{sec:conclusions}
We have analyzed the influence of polarization of spin environment on spin echo signal of the NV center qubit based on either $m\! =\! 0,1$ or $m\! =\! \pm 1$ levels, referring to these cases as the biased and unbiased qubit--environment coupling, respectively. We have considered the range of magnetic fields ($B_0\! =\! \unit{15.6}{\milli\tesla}$) and time scales (the evolution duration up to $\approx \unit{50}{\micro\second}$) in which the impact of the intrabath interactions on the course of qubit's dynamics is negligible, and the contribution from each nuclear spin can be treated independently. 
For an unpolarized environment (nuclei in a maximally mixed state), the Bloch vector of the qubit at the end of the echo sequence is parallel to its initial direction, and the coherence function is purely real. Exact calculations show then when nuclei located up to $\unit{2}{\nano\meter}$ from the qubit have polarization of the order of $50 \%$, the echoed coherence acquires a nontrivially time-dependent phase, with an amplitude $\sim \pi$ in the case of biased coupling---i.e.,~the most often investigated NV center qubit based on $m=\! 0$ and $1$ levels. For the qubit with unbiased coupling, this phase is very much smaller. This qualitative difference between the phase of the echo signal for the two kinds of couplings can be understood when one analyzes the dephasing in Gaussian approximation. An analytical, and physically transparent, expression for the phase shift was obtained for a strictly Gaussian environment in \cite{Paz_PRA17} (recounted here in Sec.~\ref{sec:Gaussian}), and it is found to be proportional to the bias. Consequently, when Gaussian approximation is valid, one should expect the above coupling-dependence, as the finite value of phase in the unbiased coupling case can only come from non-Gaussian corrections.   

We have analyzed the conditions for qubit--nucleus couplings and the evolution duration that allow one to claim that the Gaussian approximation to the NV center dephasing is in qualitative or quantitative agreement with an exact result. We have also explained the distinction between weak-coupling and Gaussian approximations---the former only requires that the influence from each individual nuclei is weak, while the latter also needs large number of nuclei. 
The necessary condition for applicability of either of these approximations is for the $\approx \unit{0.5}{\nano\meter}$ neighborhood of the NV center to be devoid of any spinful nuclei. The coupling of such a nearby nucleus with the center would be strong enough to dominate the echo dynamics. Therefore, assuming the natural concentration of ${}^{13}\mathrm{C}$ isotope in the diamond, around $50\%$ of NV centers satisfy this condition. 
The phase shift of the echo signal can be then very well described using the weak coupling approximation
If we remove also the moderately coupled nuclei located with $\unit{1}{\nano\meter}$ distance from the center, the Gaussian approximation becomes almost exact and the phase shift of unbiased qubit becomes negligible.

The main practical consequence of these results is the following: if dynamic nuclear polarization is intentionally created, or if one suspects that some polarization could have been unintentionally generated, one should measure both components of the echoed Bloch vector of the qubit (both real and imaginary part of the coherence). Measuring only the component along the initial vector can lead to apparent inconsistencies, especially when one uses theoretical model for dephasing in which all the above physics is neglected, see e.g.~\cite{Lang_PRA19}. 

Furthermore, if the presence of nuclear polarization is unclear, or its magnitude is not well characterized, one can use the measurement of phase shift of the echo signal to gain information on this polarization. In Gaussian regime, the bias-induced phase shift depends linearly on the polarization of the nuclei within its immediate vicinity ($\sim \unit{2}{\nano\meter}$). Therefore, the measurement of the phase of the echoed coherence, most conveniently accessed by measurement of the Bloch vector component perpendicular to the direction of the initial state, turns the qubit into a local probe of nuclear polarization.

A careful analysis of the phase of echoed coherence in the unbiased coupling case allows for checking to what extent the decoherence can be described as Gaussian: if it is,  the phase shift can only appear when the coupling is biased. Therefore,s a measurement of a finite phase of the echo signal of the qubit with unbiased coupling proves the non-Gaussian character of the environment. 

As discussed in \cite{Paz_PRA17}, the appearance of nonzero phase within the Gaussian approximation, is an unambiguous signature of quantum character of environmental noise. To be more precise, it means that it is, in general, impossible to simulate the dephasing by replacing the coupling to the real environment by the coupling to an external stochastic field. Let us, however, stress again that this holds true only for biased coupling $\hat H_\mathrm{int} = \lambda(\eta \hat{\mathds{1}}+\hat{\sigma}_z)\otimes\hat V/2$; the stochastic modeling of environmental influence with Gaussian noise is always possible for unbiased coupling $\hat H_\mathrm{int} = \lambda\, \hat\sigma_z\otimes \hat V/2$. In other words, the ``classical vs quantum'' nature of environmental noise is {\it relative} to the form of the qubit--environment coupling. We leave a careful examination of this issue for future research. It should stressed that a non-Gaussian external phase noise can lead to appearance of nonzero nontrivial phase shift in echo of dynamical decoupling signal, see e.g.~\cite{Cywinski_PRA14,Sung_NC19,Norris_PRL16}, so before drawing conclusions on the quantum character of the environmental influence on the qubit from the presence of the phase shift, one should carefully check if there is a nonzero Gaussian (i.e.~proportional to $\lambda^2$) contribution to this shift.
Let us also remark that another witness of quantum character of environmental influence (and also of generation of qubit--environment entanglement during dephasing) was recently described in \cite{Roszak_PRA19}, and its effectiveness for NV center qubit interacting with partially polarized nuclear environment was investigated. 

Finally, let us discuss the applicability of the above-discussed physics to the other kinds of qubits. The biased coupling, which is necessary for appearance of significant phase shift related to the ``quantum'' noise, arises for any qubit, for which only one of its states is endowed with dipole moment allowing for coupling to external field. The NV center qubit based on $m\! =\! 0$ (zero magnetic moment) and $m\! =\! 1$ (finite magnetic moment) states of the spin-1 entity, is only one example. The same coupling describes an excitonic qubit in quantum dots ($|0\rangle$ state corresponding to lack of exciton, $|1\rangle$ state to its presence) and its coupling to phonons \cite{Krummheuer_PRB02,Roszak_PLA06,Roszak_PRA06,Hodgson_PRB08,Paz_PRA17}, and for a singlet--triplet spin qubit in a double quantum dot \cite{Ramon_PRB12,Shulman_Science12} that is affected by the charge noise. While phonon baths in thermal equilibrium are exactly Gaussian (provided that anharmonicity of the lattice vibrations is negligible), environments composed of charge fluctuators \cite{Ramon_PRB12,Szankowski_JPCM17} are less obviously Gaussian, but there are regimes of environment sizes, timescales and qubit--environment couplings in which they can be treated as such to a good approximation \cite{Szankowski_JPCM17,Ramon_PRB15,Kwiatkowski_PRB18}. The conditions under which an observable phase shift of an echo signal of a singlet--triplet qubit can appear remain to be investigated.

\section*{Acknowledgements}
We thank Fattah Sakuldee and Jan Krzywda for discussions. 
This work is supported by funds of Polish National Science Center (NCN), Grant no.~2015/19/B/ST3/03152.

\appendix
\section{Dynamical decoupling control sequence}\label{sec:appendix:filterfunc}
The density matrix of the composite qubit$+$environment system at time $t$ is given by
\begin{align}
\hat\rho_{QE}(t) = \hat U_t\hat\rho_Q(0)\otimes\hat\rho_E(0)\hat U^\dagger_t, 
\end{align}
where the unitary evolution operator is given by standard time-order exponential,
\begin{align}
\hat U_t = \mathcal{T}e^{-i \int_0^t d\tau\left(\hat H_Q + \hat H_E + \hat H_\mathrm{int} + \hat H_\mathrm{ctr}(\tau)\right)}
\end{align}
with the total Hamiltonian of $Q+E$ system consists of the free environmental Hamiltonian $\hat H_E$, free qubit Hamiltonian $\hat H_Q \propto \hat\sigma_z$, the interaction $\hat H_\mathrm{int}  = \lambda(\eta \hat{\mathds{1}} + \hat\sigma_z)\otimes\hat V/2$, and the time dependent control Hamiltonian $\hat H_\mathrm{ctr}$ that is responsible for the application $\pi$ pulses in a sequence defined by the pulse timings $\{ \tau_0~=~0, \tau_1, \tau_2,\ldots, \tau_n = t\}$ ($\tau_0<\tau_1<\ldots<\tau_n$). Assuming that the pulses are perfectly accrued and effectively instantaneous, we can approximate the action of the evolution operator with
\begin{align}
\nonumber
\hat U_t &\approx
	\hat U_\mathrm{ctr}\hat U_0(t,\tau_{n-1})\ldots\hat U_\mathrm{ctr}\hat U_0(\tau_2,\tau_1)\hat U_\mathrm{ctr}\hat U_0(\tau_1,0)\\
&= (-i)^n \hat\sigma_x\hat U_0(t,\tau_{n-1})\ldots\hat\sigma_x\hat U_0(\tau_2,\tau_1)\hat\sigma_x\hat U_0(\tau_1,0)
\end{align}
where $\hat U_\mathrm{ctr} = {\exp}(i \pi \hat \sigma_x /2 ) = -i \hat\sigma_x$, and $\hat U_0(t_i,t_f) = {\exp}\{-i(t_f-t_i)[\hat H_E + \hat H_Q + \hat H_\mathrm{int}]\}$.

For any function $F$ of Pauli matrix $\hat\sigma_z$ we have the following relation
\begin{align}
\hat\sigma_x F(\hat\sigma_z)\hat\sigma_x = F(-\hat\sigma_z).
\end{align}
from which we get that
\begin{align}
\nonumber
\hat U_t &= e^{-i(t-\tau_{n-1})(-\hat H_Q + \hat H_E +\frac{\lambda}{2}(\eta\hat{\mathds{1}} + (-1)^n\hat\sigma_z)\hat V)}\ldots\\
\nonumber
&\phantom{=}\times
	 e^{-i(\tau_2-\tau_1)(-\hat H_Q+\hat H_E + \frac{\lambda}{2}(\eta\hat{\mathds{1}} - \hat \sigma_z)\hat V)}\\
\nonumber
&\phantom{=}\times
	e^{-i\tau_1(\hat H_Q+\hat H_E + \frac{\lambda}{2}(\eta\hat{\mathds{1}}+ \hat \sigma_z)\otimes\hat V)}\\
&=\mathcal{T} e^{-i \int_0^t d\tau\left(\hat H_E +f(\tau)\hat H_Q + \frac{\lambda}{2}(\eta\hat{\mathds{1}} + f(\tau)\hat\sigma_z)\otimes\hat V\right)}
\end{align}
where we encounter the time-domain filter function $f(\tau)$, defined as
\begin{align}
f(\tau) = \sum_{k=0}^{n-1}(-1)^k\Theta(\tau_{k+1} - \tau)\Theta(\tau - \tau_k).
\end{align}
We assume that the dynamical decoupling sequence is balanced, i.e.,~$\int_0^t d\tau f(\tau) = 0$, so that its application removes the trivial phase due to static energy splitting of the qubit.

After switching to interaction picture, in which $\hat V(\tau) = {\exp}(i\tau \hat H_E)\hat V {\exp}(-i\tau\hat H_E)$, we arrive at:
\begin{align}
\hat U_t = e^{-i t\hat H_E}\mathcal{T}e^{-\frac{i\lambda}{2} \int_0^t d\tau (\eta\hat{\mathds{1}} + f(\tau)\hat\sigma_z)\otimes\hat V(\tau)}
	=e^{-i t\hat H_E} \hat U(t|f).
\end{align}
Now we calculate the coherence function
\begin{align}
\nonumber
W(t) &=\frac{\mathrm{Tr}\left(\hat\sigma_{-}\otimes\hat{\mathds{1}}\left(\hat U_t \hat\rho_Q(0)\otimes\hat\rho_E(0)\hat U_t^\dagger\right)\right)}{\rho_{{+}{-}}(0)}\\
&=\frac{\mathrm{Tr}\left(\hat\sigma_{-}\otimes\hat{\mathds{1}}\left(\hat U(t|f) \hat\rho_Q(0)\otimes\hat\rho_E(0)\hat U^\dagger(t|f)\right)\right)}{\rho_{{+}{-}}(0)},
\end{align}
where the environment evolution operators $\exp(-i t\hat H_E)$ simplify due to cyclic property of the trace.



\begin{thebibliography}{51}%
\makeatletter
\providecommand \@ifxundefined [1]{%
 \@ifx{#1\undefined}
}%
\providecommand \@ifnum [1]{%
 \ifnum #1\expandafter \@firstoftwo
 \else \expandafter \@secondoftwo
 \fi
}%
\providecommand \@ifx [1]{%
 \ifx #1\expandafter \@firstoftwo
 \else \expandafter \@secondoftwo
 \fi
}%
\providecommand \natexlab [1]{#1}%
\providecommand \enquote  [1]{``#1''}%
\providecommand \bibnamefont  [1]{#1}%
\providecommand \bibfnamefont [1]{#1}%
\providecommand \citenamefont [1]{#1}%
\providecommand \href@noop [0]{\@secondoftwo}%
\providecommand \href [0]{\begingroup \@sanitize@url \@href}%
\providecommand \@href[1]{\@@startlink{#1}\@@href}%
\providecommand \@@href[1]{\endgroup#1\@@endlink}%
\providecommand \@sanitize@url [0]{\catcode `\\12\catcode `\$12\catcode
  `\&12\catcode `\#12\catcode `\^12\catcode `\_12\catcode `\%12\relax}%
\providecommand \@@startlink[1]{}%
\providecommand \@@endlink[0]{}%
\providecommand \url  [0]{\begingroup\@sanitize@url \@url }%
\providecommand \@url [1]{\endgroup\@href {#1}{\urlprefix }}%
\providecommand \urlprefix  [0]{URL }%
\providecommand \Eprint [0]{\href }%
\providecommand \doibase [0]{http://dx.doi.org/}%
\providecommand \selectlanguage [0]{\@gobble}%
\providecommand \bibinfo  [0]{\@secondoftwo}%
\providecommand \bibfield  [0]{\@secondoftwo}%
\providecommand \translation [1]{[#1]}%
\providecommand \BibitemOpen [0]{}%
\providecommand \bibitemStop [0]{}%
\providecommand \bibitemNoStop [0]{.\EOS\space}%
\providecommand \EOS [0]{\spacefactor3000\relax}%
\providecommand \BibitemShut  [1]{\csname bibitem#1\endcsname}%
\let\auto@bib@innerbib\@empty
\bibitem [{\citenamefont {Degen}\ \emph {et~al.}(2017)\citenamefont {Degen},
  \citenamefont {Reinhard},\ and\ \citenamefont {Cappellaro}}]{Degen_RMP17}%
  \BibitemOpen
  \bibfield  {author} {\bibinfo {author} {\bibfnamefont {C.~L.}\ \bibnamefont
  {Degen}}, \bibinfo {author} {\bibfnamefont {F.}~\bibnamefont {Reinhard}}, \
  and\ \bibinfo {author} {\bibfnamefont {P.}~\bibnamefont {Cappellaro}},\
  }\bibfield  {title} {\enquote {\bibinfo {title} {Quantum sensing},}\ }\href
  {\doibase 10.1103/RevModPhys.89.035002} {\bibfield  {journal} {\bibinfo
  {journal} {Rev. Mod. Phys.}\ }\textbf {\bibinfo {volume} {89}},\ \bibinfo
  {pages} {035002} (\bibinfo {year} {2017})}\BibitemShut {NoStop}%
\bibitem [{\citenamefont {Sza\'nkowski}\ \emph {et~al.}(2017)\citenamefont
  {Sza\'nkowski}, \citenamefont {Ramon}, \citenamefont {Krzywda}, \citenamefont
  {Kwiatkowski},\ and\ \citenamefont {Cywi\'nski}}]{Szankowski_JPCM17}%
  \BibitemOpen
  \bibfield  {author} {\bibinfo {author} {\bibfnamefont {P.}~\bibnamefont
  {Sza\'nkowski}}, \bibinfo {author} {\bibfnamefont {G.}~\bibnamefont {Ramon}},
  \bibinfo {author} {\bibfnamefont {J.}~\bibnamefont {Krzywda}}, \bibinfo
  {author} {\bibfnamefont {D.}~\bibnamefont {Kwiatkowski}}, \ and\ \bibinfo
  {author} {\bibfnamefont {{\L}.}~\bibnamefont {Cywi\'nski}},\ }\bibfield
  {title} {\enquote {\bibinfo {title} {Environmental noise spectroscopy with
  qubits subjected to dynamical decoupling},}\ }\href {\doibase
  10.1088/1361-648X/aa7648} {\bibfield  {journal} {\bibinfo  {journal} {J.
  Phys.:Condens. Matter}\ }\textbf {\bibinfo {volume} {29}},\ \bibinfo {pages}
  {333001} (\bibinfo {year} {2017})}\BibitemShut {NoStop}%
\bibitem [{\citenamefont {Cywi{\'n}ski}\ \emph {et~al.}(2008)\citenamefont
  {Cywi{\'n}ski}, \citenamefont {Lutchyn}, \citenamefont {Nave},\ and\
  \citenamefont {{Das Sarma}}}]{Cywinski_PRB08}%
  \BibitemOpen
  \bibfield  {author} {\bibinfo {author} {\bibfnamefont {{\L}ukasz}\
  \bibnamefont {Cywi{\'n}ski}}, \bibinfo {author} {\bibfnamefont {Roman~M.}\
  \bibnamefont {Lutchyn}}, \bibinfo {author} {\bibfnamefont {Cody~P.}\
  \bibnamefont {Nave}}, \ and\ \bibinfo {author} {\bibfnamefont
  {S.}~\bibnamefont {{Das Sarma}}},\ }\bibfield  {title} {\enquote {\bibinfo
  {title} {How to enhance dephasing time in superconducting qubits},}\ }\href
  {\doibase 10.1103/PhysRevB.77.174509} {\bibfield  {journal} {\bibinfo
  {journal} {Phys.\ Rev.\ B}\ }\textbf {\bibinfo {volume} {77}},\ \bibinfo
  {pages} {174509} (\bibinfo {year} {2008})}\BibitemShut {NoStop}%
\bibitem [{\citenamefont {Biercuk}\ \emph {et~al.}(2011)\citenamefont
  {Biercuk}, \citenamefont {Doherty},\ and\ \citenamefont
  {Uys}}]{Biercuk_JPB11}%
  \BibitemOpen
  \bibfield  {author} {\bibinfo {author} {\bibfnamefont {M.~J.}\ \bibnamefont
  {Biercuk}}, \bibinfo {author} {\bibfnamefont {A.~C.}\ \bibnamefont
  {Doherty}}, \ and\ \bibinfo {author} {\bibfnamefont {H.}~\bibnamefont
  {Uys}},\ }\bibfield  {title} {\enquote {\bibinfo {title} {Dynamical
  decoupling sequence construction as a filter-design problem},}\ }\href
  {\doibase 10.1088/0953-4075/44/15/154002} {\bibfield  {journal} {\bibinfo
  {journal} {J.~Phys.~B: At.~Mol.~Opt.~Phys.}\ }\textbf {\bibinfo {volume}
  {44}},\ \bibinfo {pages} {154002} (\bibinfo {year} {2011})}\BibitemShut
  {NoStop}%
\bibitem [{\citenamefont {Paz-Silva}\ \emph {et~al.}(2017)\citenamefont
  {Paz-Silva}, \citenamefont {Norris},\ and\ \citenamefont
  {Viola}}]{Paz_PRA17}%
  \BibitemOpen
  \bibfield  {author} {\bibinfo {author} {\bibfnamefont {Gerardo~A.}\
  \bibnamefont {Paz-Silva}}, \bibinfo {author} {\bibfnamefont {Leigh~M.}\
  \bibnamefont {Norris}}, \ and\ \bibinfo {author} {\bibfnamefont {Lorenza}\
  \bibnamefont {Viola}},\ }\bibfield  {title} {\enquote {\bibinfo {title}
  {Multiqubit spectroscopy of gaussian quantum noise},}\ }\href {\doibase
  10.1103/PhysRevA.95.022121} {\bibfield  {journal} {\bibinfo  {journal} {Phys.
  Rev. A}\ }\textbf {\bibinfo {volume} {95}},\ \bibinfo {pages} {022121}
  (\bibinfo {year} {2017})}\BibitemShut {NoStop}%
\bibitem [{\citenamefont {Dobrovitski}\ \emph {et~al.}(2013)\citenamefont
  {Dobrovitski}, \citenamefont {Fuchs}, \citenamefont {Falk}, \citenamefont
  {Santori},\ and\ \citenamefont {Awschalom}}]{Dobrovitski_ARCMP13}%
  \BibitemOpen
  \bibfield  {author} {\bibinfo {author} {\bibfnamefont {V.~V.}\ \bibnamefont
  {Dobrovitski}}, \bibinfo {author} {\bibfnamefont {G.~D.}\ \bibnamefont
  {Fuchs}}, \bibinfo {author} {\bibfnamefont {A.~L.}\ \bibnamefont {Falk}},
  \bibinfo {author} {\bibfnamefont {C.}~\bibnamefont {Santori}}, \ and\
  \bibinfo {author} {\bibfnamefont {D.~D.}\ \bibnamefont {Awschalom}},\
  }\bibfield  {title} {\enquote {\bibinfo {title} {Quantum control over single
  spins in diamond},}\ }\href {\doibase
  10.1146/annurev-conmatphys-030212-184238} {\bibfield  {journal} {\bibinfo
  {journal} {Ann. Rev. Cond. Mat. Phys.}\ }\textbf {\bibinfo {volume} {4}},\
  \bibinfo {pages} {23} (\bibinfo {year} {2013})}\BibitemShut {NoStop}%
\bibitem [{\citenamefont {Rondin}\ \emph {et~al.}(2014)\citenamefont {Rondin},
  \citenamefont {Tetienne}, \citenamefont {Hingant}, \citenamefont {Roch},
  \citenamefont {Maletinsky},\ and\ \citenamefont {Jacques}}]{Rondin_RPP14}%
  \BibitemOpen
  \bibfield  {author} {\bibinfo {author} {\bibfnamefont {L.}~\bibnamefont
  {Rondin}}, \bibinfo {author} {\bibfnamefont {J.-P.}\ \bibnamefont
  {Tetienne}}, \bibinfo {author} {\bibfnamefont {T.}~\bibnamefont {Hingant}},
  \bibinfo {author} {\bibfnamefont {J.-F.}\ \bibnamefont {Roch}}, \bibinfo
  {author} {\bibfnamefont {P.}~\bibnamefont {Maletinsky}}, \ and\ \bibinfo
  {author} {\bibfnamefont {V.}~\bibnamefont {Jacques}},\ }\bibfield  {title}
  {\enquote {\bibinfo {title} {Magnetometry with nitrogen-vacancy defects in
  diamond},}\ }\href {\doibase 10.1088/0034-4885/77/5/056503} {\bibfield
  {journal} {\bibinfo  {journal} {Rep.~Prog.~Phys.}\ }\textbf {\bibinfo
  {volume} {77}},\ \bibinfo {pages} {056503} (\bibinfo {year}
  {2014})}\BibitemShut {NoStop}%
\bibitem [{\citenamefont {Coish}\ and\ \citenamefont
  {Baugh}(2009)}]{Coish_PSSB09}%
  \BibitemOpen
  \bibfield  {author} {\bibinfo {author} {\bibfnamefont {W.~A.}\ \bibnamefont
  {Coish}}\ and\ \bibinfo {author} {\bibfnamefont {J.}~\bibnamefont {Baugh}},\
  }\bibfield  {title} {\enquote {\bibinfo {title} {Nuclear spins in
  nanostructures},}\ }\href {\doibase 10.1002/pssb.200945229} {\bibfield
  {journal} {\bibinfo  {journal} {Phys. Status Solidi B}\ }\textbf {\bibinfo
  {volume} {246}},\ \bibinfo {pages} {2203} (\bibinfo {year}
  {2009})}\BibitemShut {NoStop}%
\bibitem [{\citenamefont {Cywi{\'n}ski}(2011)}]{Cywinski_APPA11}%
  \BibitemOpen
  \bibfield  {author} {\bibinfo {author} {\bibfnamefont {{\L}ukasz}\
  \bibnamefont {Cywi{\'n}ski}},\ }\bibfield  {title} {\enquote {\bibinfo
  {title} {Dephasing of electron spin qubits due to their interaction with
  nuclei in quantum dots},}\ }\href {\doibase 10.12693/APhysPolA.119.576}
  {\bibfield  {journal} {\bibinfo  {journal} {Acta Phys.~Pol.~A}\ }\textbf
  {\bibinfo {volume} {119}},\ \bibinfo {pages} {576} (\bibinfo {year}
  {2011})}\BibitemShut {NoStop}%
\bibitem [{\citenamefont {Chekhovich}\ \emph {et~al.}(2013)\citenamefont
  {Chekhovich}, \citenamefont {Makhonin}, \citenamefont {Tartakovskii},
  \citenamefont {Yacoby}, \citenamefont {Bluhm}, \citenamefont {Nowack},\ and\
  \citenamefont {Vandersypen}}]{Chekhovich_NM13}%
  \BibitemOpen
  \bibfield  {author} {\bibinfo {author} {\bibfnamefont {E.~A.}\ \bibnamefont
  {Chekhovich}}, \bibinfo {author} {\bibfnamefont {M.~N.}\ \bibnamefont
  {Makhonin}}, \bibinfo {author} {\bibfnamefont {A.~I.}\ \bibnamefont
  {Tartakovskii}}, \bibinfo {author} {\bibfnamefont {A.}~\bibnamefont
  {Yacoby}}, \bibinfo {author} {\bibfnamefont {H.}~\bibnamefont {Bluhm}},
  \bibinfo {author} {\bibfnamefont {K.~C.}\ \bibnamefont {Nowack}}, \ and\
  \bibinfo {author} {\bibfnamefont {L.~M.~K.}\ \bibnamefont {Vandersypen}},\
  }\bibfield  {title} {\enquote {\bibinfo {title} {Nuclear spin effects in
  semiconductor quantum dots},}\ }\href {\doibase 10.1038/nmat3652} {\bibfield
  {journal} {\bibinfo  {journal} {Nature Materials}\ }\textbf {\bibinfo
  {volume} {12}},\ \bibinfo {pages} {494} (\bibinfo {year} {2013})}\BibitemShut
  {NoStop}%
\bibitem [{\citenamefont {Kwiatkowski}\ and\ \citenamefont
  {Cywi\'{n}ski}(2018)}]{Kwiatkowski_PRB18}%
  \BibitemOpen
  \bibfield  {author} {\bibinfo {author} {\bibfnamefont {D.}~\bibnamefont
  {Kwiatkowski}}\ and\ \bibinfo {author} {\bibfnamefont {\L.}\ \bibnamefont
  {Cywi\'{n}ski}},\ }\bibfield  {title} {\enquote {\bibinfo {title}
  {Decoherence of two entangled spin qubits coupled to an interacting sparse
  nuclear spin bath: Application to nitrogen vacancy centers},}\ }\href
  {\doibase 10.1103/PhysRevB.98.155202} {\bibfield  {journal} {\bibinfo
  {journal} {Phys. Rev. B}\ }\textbf {\bibinfo {volume} {98}},\ \bibinfo
  {pages} {155202} (\bibinfo {year} {2018})}\BibitemShut {NoStop}%
\bibitem [{\citenamefont {London}\ \emph {et~al.}(2013)\citenamefont {London},
  \citenamefont {Scheuer}, \citenamefont {Cai}, \citenamefont {Schwarz},
  \citenamefont {Retzker}, \citenamefont {Plenio}, \citenamefont {Katagiri},
  \citenamefont {Teraji}, \citenamefont {Koizumi}, \citenamefont {Isoya},
  \citenamefont {Fischer}, \citenamefont {McGuinness}, \citenamefont
  {Naydenov},\ and\ \citenamefont {Jelezko}}]{London2013}%
  \BibitemOpen
  \bibfield  {author} {\bibinfo {author} {\bibfnamefont {P.}~\bibnamefont
  {London}}, \bibinfo {author} {\bibfnamefont {J.}~\bibnamefont {Scheuer}},
  \bibinfo {author} {\bibfnamefont {J.~M.}\ \bibnamefont {Cai}}, \bibinfo
  {author} {\bibfnamefont {I.}~\bibnamefont {Schwarz}}, \bibinfo {author}
  {\bibfnamefont {A.}~\bibnamefont {Retzker}}, \bibinfo {author} {\bibfnamefont
  {M.~B.}\ \bibnamefont {Plenio}}, \bibinfo {author} {\bibfnamefont
  {M.}~\bibnamefont {Katagiri}}, \bibinfo {author} {\bibfnamefont
  {T.}~\bibnamefont {Teraji}}, \bibinfo {author} {\bibfnamefont
  {S.}~\bibnamefont {Koizumi}}, \bibinfo {author} {\bibfnamefont
  {J.}~\bibnamefont {Isoya}}, \bibinfo {author} {\bibfnamefont
  {R.}~\bibnamefont {Fischer}}, \bibinfo {author} {\bibfnamefont {L.~P.}\
  \bibnamefont {McGuinness}}, \bibinfo {author} {\bibfnamefont
  {B.}~\bibnamefont {Naydenov}}, \ and\ \bibinfo {author} {\bibfnamefont
  {F.}~\bibnamefont {Jelezko}},\ }\bibfield  {title} {\enquote {\bibinfo
  {title} {{Detecting and polarizing nuclear spins with double resonance on a
  single electron spin}},}\ }\href {\doibase 10.1103/PhysRevLett.111.067601}
  {\bibfield  {journal} {\bibinfo  {journal} {Phys.~Rev.~Lett.}\ }\textbf
  {\bibinfo {volume} {111}},\ \bibinfo {pages} {067601} (\bibinfo {year}
  {2013})}\BibitemShut {NoStop}%
\bibitem [{\citenamefont {Fischer}\ \emph {et~al.}(2013)\citenamefont
  {Fischer}, \citenamefont {Bretschneider}, \citenamefont {London},
  \citenamefont {Budker}, \citenamefont {Gershoni},\ and\ \citenamefont
  {Frydman}}]{Fischer2013a}%
  \BibitemOpen
  \bibfield  {author} {\bibinfo {author} {\bibfnamefont {Ran}\ \bibnamefont
  {Fischer}}, \bibinfo {author} {\bibfnamefont {Christian~O.}\ \bibnamefont
  {Bretschneider}}, \bibinfo {author} {\bibfnamefont {Paz}\ \bibnamefont
  {London}}, \bibinfo {author} {\bibfnamefont {Dmitry}\ \bibnamefont {Budker}},
  \bibinfo {author} {\bibfnamefont {David}\ \bibnamefont {Gershoni}}, \ and\
  \bibinfo {author} {\bibfnamefont {Lucio}\ \bibnamefont {Frydman}},\
  }\bibfield  {title} {\enquote {\bibinfo {title} {{Bulk nuclear polarization
  enhanced at room temperature by optical pumping}},}\ }\href {\doibase
  10.1103/PhysRevLett.111.057601} {\bibfield  {journal} {\bibinfo  {journal}
  {Physical Review Letters}\ }\textbf {\bibinfo {volume} {111}},\ \bibinfo
  {pages} {057601} (\bibinfo {year} {2013})}\BibitemShut {NoStop}%
\bibitem [{\citenamefont {Pagliero}\ \emph {et~al.}(2018)\citenamefont
  {Pagliero}, \citenamefont {Rao}, \citenamefont {Zangara}, \citenamefont
  {Dhomkar}, \citenamefont {Wong}, \citenamefont {Abril}, \citenamefont
  {Aslam}, \citenamefont {Parker}, \citenamefont {King}, \citenamefont
  {Avalos}, \citenamefont {Ajoy}, \citenamefont {Wrachtrup}, \citenamefont
  {Pines},\ and\ \citenamefont {Meriles}}]{Pagliero2018}%
  \BibitemOpen
  \bibfield  {author} {\bibinfo {author} {\bibfnamefont {Daniela}\ \bibnamefont
  {Pagliero}}, \bibinfo {author} {\bibfnamefont {K.~R.~Koteswara}\ \bibnamefont
  {Rao}}, \bibinfo {author} {\bibfnamefont {Pablo~R.}\ \bibnamefont {Zangara}},
  \bibinfo {author} {\bibfnamefont {Siddharth}\ \bibnamefont {Dhomkar}},
  \bibinfo {author} {\bibfnamefont {Henry~H.}\ \bibnamefont {Wong}}, \bibinfo
  {author} {\bibfnamefont {Andrea}\ \bibnamefont {Abril}}, \bibinfo {author}
  {\bibfnamefont {Nabeel}\ \bibnamefont {Aslam}}, \bibinfo {author}
  {\bibfnamefont {Anna}\ \bibnamefont {Parker}}, \bibinfo {author}
  {\bibfnamefont {Jonathan}\ \bibnamefont {King}}, \bibinfo {author}
  {\bibfnamefont {Claudia~E.}\ \bibnamefont {Avalos}}, \bibinfo {author}
  {\bibfnamefont {Ashok}\ \bibnamefont {Ajoy}}, \bibinfo {author}
  {\bibfnamefont {Joerg}\ \bibnamefont {Wrachtrup}}, \bibinfo {author}
  {\bibfnamefont {Alexander}\ \bibnamefont {Pines}}, \ and\ \bibinfo {author}
  {\bibfnamefont {Carlos~A.}\ \bibnamefont {Meriles}},\ }\bibfield  {title}
  {\enquote {\bibinfo {title} {Multispin-assisted optical pumping of bulk
  $^{13}\mathrm{C}$ nuclear spin polarization in diamond},}\ }\href {\doibase
  10.1103/PhysRevB.97.024422} {\bibfield  {journal} {\bibinfo  {journal} {Phys.
  Rev. B}\ }\textbf {\bibinfo {volume} {97}},\ \bibinfo {pages} {024422}
  (\bibinfo {year} {2018})}\BibitemShut {NoStop}%
\bibitem [{\citenamefont {Wunderlich}\ \emph {et~al.}(2017)\citenamefont
  {Wunderlich}, \citenamefont {Kohlrautz}, \citenamefont {Abel}, \citenamefont
  {Haase},\ and\ \citenamefont {Meijer}}]{Wunderlich2017}%
  \BibitemOpen
  \bibfield  {author} {\bibinfo {author} {\bibfnamefont {Ralf}\ \bibnamefont
  {Wunderlich}}, \bibinfo {author} {\bibfnamefont {Jonas}\ \bibnamefont
  {Kohlrautz}}, \bibinfo {author} {\bibfnamefont {Bernd}\ \bibnamefont {Abel}},
  \bibinfo {author} {\bibfnamefont {J{\"{u}}rgen}\ \bibnamefont {Haase}}, \
  and\ \bibinfo {author} {\bibfnamefont {Jan}\ \bibnamefont {Meijer}},\
  }\bibfield  {title} {\enquote {\bibinfo {title} {{Optically induced cross
  relaxation via nitrogen-related defects for bulk diamond C 13
  hyperpolarization}},}\ }\href {\doibase 10.1103/PhysRevB.96.220407}
  {\bibfield  {journal} {\bibinfo  {journal} {Phys.~Rev.~B}\ }\textbf {\bibinfo
  {volume} {96}},\ \bibinfo {pages} {220407(R)} (\bibinfo {year}
  {2017})}\BibitemShut {NoStop}%
\bibitem [{\citenamefont {{\'{A}}lvarez}\ \emph {et~al.}(2015)\citenamefont
  {{\'{A}}lvarez}, \citenamefont {Bretschneider}, \citenamefont {Fischer},
  \citenamefont {London}, \citenamefont {Kanda}, \citenamefont {Onoda},
  \citenamefont {Isoya}, \citenamefont {Gershoni},\ and\ \citenamefont
  {Frydman}}]{Alvarez2015}%
  \BibitemOpen
  \bibfield  {author} {\bibinfo {author} {\bibfnamefont {Gonzalo~A.}\
  \bibnamefont {{\'{A}}lvarez}}, \bibinfo {author} {\bibfnamefont
  {Christian~O.}\ \bibnamefont {Bretschneider}}, \bibinfo {author}
  {\bibfnamefont {Ran}\ \bibnamefont {Fischer}}, \bibinfo {author}
  {\bibfnamefont {Paz}\ \bibnamefont {London}}, \bibinfo {author}
  {\bibfnamefont {Hisao}\ \bibnamefont {Kanda}}, \bibinfo {author}
  {\bibfnamefont {Shinobu}\ \bibnamefont {Onoda}}, \bibinfo {author}
  {\bibfnamefont {Junichi}\ \bibnamefont {Isoya}}, \bibinfo {author}
  {\bibfnamefont {David}\ \bibnamefont {Gershoni}}, \ and\ \bibinfo {author}
  {\bibfnamefont {Lucio}\ \bibnamefont {Frydman}},\ }\bibfield  {title}
  {\enquote {\bibinfo {title} {{Local and bulk 13C hyperpolarization in
  nitrogen-vacancy-centred diamonds at variable fields and orientations}},}\
  }\href {\doibase 10.1038/ncomms9456} {\bibfield  {journal} {\bibinfo
  {journal} {Nature Communications}\ }\textbf {\bibinfo {volume} {6}},\
  \bibinfo {pages} {8456} (\bibinfo {year} {2015})},\ \Eprint
  {http://arxiv.org/abs/1412.8635} {1412.8635} \BibitemShut {NoStop}%
\bibitem [{\citenamefont {King}\ \emph {et~al.}(2015)\citenamefont {King},
  \citenamefont {Jeong}, \citenamefont {Vassiliou}, \citenamefont {Shin},
  \citenamefont {Page}, \citenamefont {Avalos}, \citenamefont {Wang},\ and\
  \citenamefont {Pines}}]{King2015}%
  \BibitemOpen
  \bibfield  {author} {\bibinfo {author} {\bibfnamefont {Jonathan~P.}\
  \bibnamefont {King}}, \bibinfo {author} {\bibfnamefont {Keunhong}\
  \bibnamefont {Jeong}}, \bibinfo {author} {\bibfnamefont {Christophoros~C.}\
  \bibnamefont {Vassiliou}}, \bibinfo {author} {\bibfnamefont {Chang~S.}\
  \bibnamefont {Shin}}, \bibinfo {author} {\bibfnamefont {Ralph~H.}\
  \bibnamefont {Page}}, \bibinfo {author} {\bibfnamefont {Claudia~E.}\
  \bibnamefont {Avalos}}, \bibinfo {author} {\bibfnamefont {Hai-Jing}\
  \bibnamefont {Wang}}, \ and\ \bibinfo {author} {\bibfnamefont {Alexander}\
  \bibnamefont {Pines}},\ }\bibfield  {title} {\enquote {\bibinfo {title}
  {{Room-temperature in situ nuclear spin hyperpolarization from optically
  pumped nitrogen vacancy centres in diamond}},}\ }\href {\doibase
  10.1038/ncomms9965} {\bibfield  {journal} {\bibinfo  {journal} {Nature
  Communications}\ }\textbf {\bibinfo {volume} {6}},\ \bibinfo {pages} {8965}
  (\bibinfo {year} {2015})}\BibitemShut {NoStop}%
\bibitem [{\citenamefont {Scheuer}\ \emph {et~al.}(2017)\citenamefont
  {Scheuer}, \citenamefont {Schwartz}, \citenamefont {M{\"{u}}ller},
  \citenamefont {Chen}, \citenamefont {Dhand}, \citenamefont {Plenio},
  \citenamefont {Naydenov},\ and\ \citenamefont {Jelezko}}]{Scheuer2017}%
  \BibitemOpen
  \bibfield  {author} {\bibinfo {author} {\bibfnamefont {Jochen}\ \bibnamefont
  {Scheuer}}, \bibinfo {author} {\bibfnamefont {Ilai}\ \bibnamefont
  {Schwartz}}, \bibinfo {author} {\bibfnamefont {Samuel}\ \bibnamefont
  {M{\"{u}}ller}}, \bibinfo {author} {\bibfnamefont {Qiong}\ \bibnamefont
  {Chen}}, \bibinfo {author} {\bibfnamefont {Ish}\ \bibnamefont {Dhand}},
  \bibinfo {author} {\bibfnamefont {Martin~B.}\ \bibnamefont {Plenio}},
  \bibinfo {author} {\bibfnamefont {Boris}\ \bibnamefont {Naydenov}}, \ and\
  \bibinfo {author} {\bibfnamefont {Fedor}\ \bibnamefont {Jelezko}},\
  }\bibfield  {title} {\enquote {\bibinfo {title} {{Robust techniques for
  polarization and detection of nuclear spin ensembles}},}\ }\href {\doibase
  10.1103/PhysRevB.96.174436} {\bibfield  {journal} {\bibinfo  {journal}
  {Physical Review B}\ }\textbf {\bibinfo {volume} {96}},\ \bibinfo {pages}
  {174436} (\bibinfo {year} {2017})},\ \Eprint
  {http://arxiv.org/abs/1706.01315} {arXiv:1706.01315} \BibitemShut {NoStop}%
\bibitem [{\citenamefont {Poggiali}\ \emph {et~al.}(2017)\citenamefont
  {Poggiali}, \citenamefont {Cappellaro},\ and\ \citenamefont
  {Fabbri}}]{Poggiali2017}%
  \BibitemOpen
  \bibfield  {author} {\bibinfo {author} {\bibfnamefont {F.}~\bibnamefont
  {Poggiali}}, \bibinfo {author} {\bibfnamefont {P.}~\bibnamefont
  {Cappellaro}}, \ and\ \bibinfo {author} {\bibfnamefont {N.}~\bibnamefont
  {Fabbri}},\ }\bibfield  {title} {\enquote {\bibinfo {title} {{Measurement of
  the excited-state transverse hyperfine coupling in NV centers via dynamic
  nuclear polarization}},}\ }\href {\doibase 10.1103/PhysRevB.95.195308}
  {\bibfield  {journal} {\bibinfo  {journal} {Physical Review B}\ }\textbf
  {\bibinfo {volume} {95}},\ \bibinfo {pages} {195308} (\bibinfo {year}
  {2017})},\ \Eprint {http://arxiv.org/abs/1612.04783} {arXiv:1612.04783}
  \BibitemShut {NoStop}%
\bibitem [{\citenamefont {Hovav}\ \emph {et~al.}(2018)\citenamefont {Hovav},
  \citenamefont {Naydenov}, \citenamefont {Jelezko},\ and\ \citenamefont
  {Bar-Gill}}]{Hovav2018}%
  \BibitemOpen
  \bibfield  {author} {\bibinfo {author} {\bibfnamefont {Y.}~\bibnamefont
  {Hovav}}, \bibinfo {author} {\bibfnamefont {B.}~\bibnamefont {Naydenov}},
  \bibinfo {author} {\bibfnamefont {F.}~\bibnamefont {Jelezko}}, \ and\
  \bibinfo {author} {\bibfnamefont {N.}~\bibnamefont {Bar-Gill}},\ }\bibfield
  {title} {\enquote {\bibinfo {title} {Low-field nuclear polarization using
  nitrogen vacancy centers in diamonds},}\ }\href {\doibase
  10.1103/PhysRevLett.120.060405} {\bibfield  {journal} {\bibinfo  {journal}
  {Phys. Rev. Lett.}\ }\textbf {\bibinfo {volume} {120}},\ \bibinfo {pages}
  {060405} (\bibinfo {year} {2018})}\BibitemShut {NoStop}%
\bibitem [{\citenamefont {Yang}\ and\ \citenamefont
  {Liu}(2008)}]{Yang_CCE_PRB08}%
  \BibitemOpen
  \bibfield  {author} {\bibinfo {author} {\bibfnamefont {Wen}\ \bibnamefont
  {Yang}}\ and\ \bibinfo {author} {\bibfnamefont {Ren-Bao}\ \bibnamefont
  {Liu}},\ }\bibfield  {title} {\enquote {\bibinfo {title} {Quantum many-body
  theory of qubit decoherence in a finite-size spin bath},}\ }\href {\doibase
  10.1103/PhysRevB.78.085315} {\bibfield  {journal} {\bibinfo  {journal}
  {Phys.\ Rev.\ B}\ }\textbf {\bibinfo {volume} {78}},\ \bibinfo {pages}
  {085315} (\bibinfo {year} {2008})}\BibitemShut {NoStop}%
\bibitem [{\citenamefont {Zhao}\ \emph {et~al.}(2012)\citenamefont {Zhao},
  \citenamefont {Ho},\ and\ \citenamefont {Liu}}]{Zhao_PRB12}%
  \BibitemOpen
  \bibfield  {author} {\bibinfo {author} {\bibfnamefont {Nan}\ \bibnamefont
  {Zhao}}, \bibinfo {author} {\bibfnamefont {Sai-Wah}\ \bibnamefont {Ho}}, \
  and\ \bibinfo {author} {\bibfnamefont {Ren-Bao}\ \bibnamefont {Liu}},\
  }\bibfield  {title} {\enquote {\bibinfo {title} {Decoherence and dynamical
  decoupling control of nitrogen vacancy center electron spins in nuclear spin
  baths},}\ }\href {\doibase 10.1103/PhysRevB.85.115303} {\bibfield  {journal}
  {\bibinfo  {journal} {Phys. Rev. B}\ }\textbf {\bibinfo {volume} {85}},\
  \bibinfo {pages} {115303} (\bibinfo {year} {2012})}\BibitemShut {NoStop}%
\bibitem [{\citenamefont {Yang}\ \emph {et~al.}(2017)\citenamefont {Yang},
  \citenamefont {Ma},\ and\ \citenamefont {Liu}}]{Yang_RPP17}%
  \BibitemOpen
  \bibfield  {author} {\bibinfo {author} {\bibfnamefont {Wen}\ \bibnamefont
  {Yang}}, \bibinfo {author} {\bibfnamefont {Wen-Long}\ \bibnamefont {Ma}}, \
  and\ \bibinfo {author} {\bibfnamefont {Ren-Bao}\ \bibnamefont {Liu}},\
  }\bibfield  {title} {\enquote {\bibinfo {title} {Quantum many-body theory for
  electron spin decoherence in nanoscale nuclear spin baths},}\ }\href
  {\doibase 10.1088/0034-4885/80/1/016001} {\bibfield  {journal} {\bibinfo
  {journal} {Rep. Prog. Phys.}\ }\textbf {\bibinfo {volume} {80}},\ \bibinfo
  {pages} {016001} (\bibinfo {year} {2017})}\BibitemShut {NoStop}%
\bibitem [{\citenamefont {Viola}\ and\ \citenamefont
  {Lloyd}(1998)}]{Viola_PRA98}%
  \BibitemOpen
  \bibfield  {author} {\bibinfo {author} {\bibfnamefont {Lorenza}\ \bibnamefont
  {Viola}}\ and\ \bibinfo {author} {\bibfnamefont {Seth}\ \bibnamefont
  {Lloyd}},\ }\bibfield  {title} {\enquote {\bibinfo {title} {Dynamical
  suppression of decoherence in two-state quantum systems},}\ }\href {\doibase
  10.1103/PhysRevA.58.2733} {\bibfield  {journal} {\bibinfo  {journal} {Phys.
  Rev. A}\ }\textbf {\bibinfo {volume} {58}},\ \bibinfo {pages} {2733}
  (\bibinfo {year} {1998})}\BibitemShut {NoStop}%
\bibitem [{\citenamefont {Suter}\ and\ \citenamefont
  {\'Alvarez}(2016)}]{Suter_RMP16}%
  \BibitemOpen
  \bibfield  {author} {\bibinfo {author} {\bibfnamefont {Dieter}\ \bibnamefont
  {Suter}}\ and\ \bibinfo {author} {\bibfnamefont {Gonzalo~A.}\ \bibnamefont
  {\'Alvarez}},\ }\bibfield  {title} {\enquote {\bibinfo {title} {Colloquium:
  Protecting quantum information against environmental noise},}\ }\href
  {\doibase 10.1103/RevModPhys.88.041001} {\bibfield  {journal} {\bibinfo
  {journal} {Rev. Mod. Phys.}\ }\textbf {\bibinfo {volume} {88}},\ \bibinfo
  {pages} {041001} (\bibinfo {year} {2016})}\BibitemShut {NoStop}%
\bibitem [{\citenamefont {Huang}\ \emph {et~al.}(2011)\citenamefont {Huang},
  \citenamefont {Kong}, \citenamefont {Zhao}, \citenamefont {Shi},
  \citenamefont {Wang}, \citenamefont {Rong}, \citenamefont {Liu},\ and\
  \citenamefont {Du}}]{Huang_NC11}%
  \BibitemOpen
  \bibfield  {author} {\bibinfo {author} {\bibfnamefont {Pu}~\bibnamefont
  {Huang}}, \bibinfo {author} {\bibfnamefont {Xi}~\bibnamefont {Kong}},
  \bibinfo {author} {\bibfnamefont {Nan}\ \bibnamefont {Zhao}}, \bibinfo
  {author} {\bibfnamefont {Fazhan}\ \bibnamefont {Shi}}, \bibinfo {author}
  {\bibfnamefont {Pengfei}\ \bibnamefont {Wang}}, \bibinfo {author}
  {\bibfnamefont {Xing}\ \bibnamefont {Rong}}, \bibinfo {author} {\bibfnamefont
  {Ren-Bao}\ \bibnamefont {Liu}}, \ and\ \bibinfo {author} {\bibfnamefont
  {Jiangfeng}\ \bibnamefont {Du}},\ }\bibfield  {title} {\enquote {\bibinfo
  {title} {Observation of anomalous decoherence effect in a quantum bath at
  room temperature},}\ }\href {\doibase 10.1038/ncomms1579} {\bibfield
  {journal} {\bibinfo  {journal} {Nat.~Communications}\ }\textbf {\bibinfo
  {volume} {2}},\ \bibinfo {pages} {570} (\bibinfo {year} {2011})}\BibitemShut
  {NoStop}%
\bibitem [{\citenamefont {Dolde}\ \emph {et~al.}(2013)\citenamefont {Dolde},
  \citenamefont {Jakobi}, \citenamefont {Naydenov}, \citenamefont {Zhao},
  \citenamefont {Pezzagna}, \citenamefont {Trautmann}, \citenamefont {Meijer},
  \citenamefont {Neumann}, \citenamefont {Jelezko},\ and\ \citenamefont
  {J.Wrachtrup}}]{Dolde_NP13}%
  \BibitemOpen
  \bibfield  {author} {\bibinfo {author} {\bibfnamefont {F.}~\bibnamefont
  {Dolde}}, \bibinfo {author} {\bibfnamefont {I.}~\bibnamefont {Jakobi}},
  \bibinfo {author} {\bibfnamefont {B.}~\bibnamefont {Naydenov}}, \bibinfo
  {author} {\bibfnamefont {N.}~\bibnamefont {Zhao}}, \bibinfo {author}
  {\bibfnamefont {S.}~\bibnamefont {Pezzagna}}, \bibinfo {author}
  {\bibfnamefont {C.}~\bibnamefont {Trautmann}}, \bibinfo {author}
  {\bibfnamefont {J.}~\bibnamefont {Meijer}}, \bibinfo {author} {\bibfnamefont
  {P.}~\bibnamefont {Neumann}}, \bibinfo {author} {\bibfnamefont
  {F.}~\bibnamefont {Jelezko}}, \ and\ \bibinfo {author} {\bibnamefont
  {J.Wrachtrup}},\ }\bibfield  {title} {\enquote {\bibinfo {title}
  {Room-temperature entanglement between single defect spins in diamond},}\
  }\href {\doibase 10.1038/nphys2545} {\bibfield  {journal} {\bibinfo
  {journal} {Nat. Phys.}\ }\textbf {\bibinfo {volume} {9}},\ \bibinfo {pages}
  {139} (\bibinfo {year} {2013})}\BibitemShut {NoStop}%
\bibitem [{\citenamefont {Staudacher}\ \emph {et~al.}(2013)\citenamefont
  {Staudacher}, \citenamefont {Shi}, \citenamefont {Pezzagna}, \citenamefont
  {Meijer}, \citenamefont {Du}, \citenamefont {Meriles}, \citenamefont
  {Reinhard},\ and\ \citenamefont {Wrachtrup}}]{Staudacher_Science13}%
  \BibitemOpen
  \bibfield  {author} {\bibinfo {author} {\bibfnamefont {T.}~\bibnamefont
  {Staudacher}}, \bibinfo {author} {\bibfnamefont {F.}~\bibnamefont {Shi}},
  \bibinfo {author} {\bibfnamefont {S.}~\bibnamefont {Pezzagna}}, \bibinfo
  {author} {\bibfnamefont {J.}~\bibnamefont {Meijer}}, \bibinfo {author}
  {\bibfnamefont {J.}~\bibnamefont {Du}}, \bibinfo {author} {\bibfnamefont
  {C.~A.}\ \bibnamefont {Meriles}}, \bibinfo {author} {\bibfnamefont
  {F.}~\bibnamefont {Reinhard}}, \ and\ \bibinfo {author} {\bibfnamefont
  {J.}~\bibnamefont {Wrachtrup}},\ }\bibfield  {title} {\enquote {\bibinfo
  {title} {Nuclear magnetic resonance spectroscopy on a (5-nanometer)$^{3}$
  sample volume},}\ }\href {\doibase 10.1126/science.1231675} {\bibfield
  {journal} {\bibinfo  {journal} {Science}\ }\textbf {\bibinfo {volume}
  {339}},\ \bibinfo {pages} {561} (\bibinfo {year} {2013})}\BibitemShut
  {NoStop}%
\bibitem [{\citenamefont {H{\"a}berle}\ \emph {et~al.}(2015)\citenamefont
  {H{\"a}berle}, \citenamefont {Schmid-Lorch}, \citenamefont {Reinhard},\ and\
  \citenamefont {Wrachtrup}}]{Haberle_NN15}%
  \BibitemOpen
  \bibfield  {author} {\bibinfo {author} {\bibfnamefont {T.}~\bibnamefont
  {H{\"a}berle}}, \bibinfo {author} {\bibfnamefont {D.}~\bibnamefont
  {Schmid-Lorch}}, \bibinfo {author} {\bibfnamefont {F.}~\bibnamefont
  {Reinhard}}, \ and\ \bibinfo {author} {\bibfnamefont {J.}~\bibnamefont
  {Wrachtrup}},\ }\bibfield  {title} {\enquote {\bibinfo {title} {Nanoscale
  nuclear magnetic imaging with chemical contrast},}\ }\href {\doibase
  10.1038/nnano.2014.299} {\bibfield  {journal} {\bibinfo  {journal} {Nature
  Nanotechnology}\ }\textbf {\bibinfo {volume} {10}},\ \bibinfo {pages} {125}
  (\bibinfo {year} {2015})}\BibitemShut {NoStop}%
\bibitem [{\citenamefont {DeVience}\ \emph {et~al.}(2015)\citenamefont
  {DeVience}, \citenamefont {Pham}, \citenamefont {Lovchinsky}, \citenamefont
  {Sushkov}, \citenamefont {Bar-Gill}, \citenamefont {Belthangady},
  \citenamefont {Casola}, \citenamefont {Corbett}, \citenamefont {Zhang},
  \citenamefont {Lukin}, \citenamefont {Park}, \citenamefont {Yacoby},\ and\
  \citenamefont {Walsworth}}]{DeVience_NN15}%
  \BibitemOpen
  \bibfield  {author} {\bibinfo {author} {\bibfnamefont {Stephen~J.}\
  \bibnamefont {DeVience}}, \bibinfo {author} {\bibfnamefont {Linh~M.}\
  \bibnamefont {Pham}}, \bibinfo {author} {\bibfnamefont {Igor}\ \bibnamefont
  {Lovchinsky}}, \bibinfo {author} {\bibfnamefont {Alexander~O.}\ \bibnamefont
  {Sushkov}}, \bibinfo {author} {\bibfnamefont {Nir}\ \bibnamefont {Bar-Gill}},
  \bibinfo {author} {\bibfnamefont {Chinmay}\ \bibnamefont {Belthangady}},
  \bibinfo {author} {\bibfnamefont {Francesco}\ \bibnamefont {Casola}},
  \bibinfo {author} {\bibfnamefont {Madeleine}\ \bibnamefont {Corbett}},
  \bibinfo {author} {\bibfnamefont {Huiliang}\ \bibnamefont {Zhang}}, \bibinfo
  {author} {\bibfnamefont {Mikhail}\ \bibnamefont {Lukin}}, \bibinfo {author}
  {\bibfnamefont {Hongkun}\ \bibnamefont {Park}}, \bibinfo {author}
  {\bibfnamefont {Amir}\ \bibnamefont {Yacoby}}, \ and\ \bibinfo {author}
  {\bibfnamefont {Ronald~L.}\ \bibnamefont {Walsworth}},\ }\bibfield  {title}
  {\enquote {\bibinfo {title} {Nanoscale nmr spectroscopy and imaging of
  multiple nuclear species},}\ }\href {\doibase 10.1038/nnano.2014.313}
  {\bibfield  {journal} {\bibinfo  {journal} {Nature Nanotechnology}\ }\textbf
  {\bibinfo {volume} {10}},\ \bibinfo {pages} {129} (\bibinfo {year}
  {2015})}\BibitemShut {NoStop}%
\bibitem [{\citenamefont {Hern\'andez-G\'omez}\ \emph
  {et~al.}(2018)\citenamefont {Hern\'andez-G\'omez}, \citenamefont {Poggiali},
  \citenamefont {Cappellaro},\ and\ \citenamefont {Fabbri}}]{Hernandez_PRB18}%
  \BibitemOpen
  \bibfield  {author} {\bibinfo {author} {\bibfnamefont {S.}~\bibnamefont
  {Hern\'andez-G\'omez}}, \bibinfo {author} {\bibfnamefont {F.}~\bibnamefont
  {Poggiali}}, \bibinfo {author} {\bibfnamefont {P.}~\bibnamefont
  {Cappellaro}}, \ and\ \bibinfo {author} {\bibfnamefont {N.}~\bibnamefont
  {Fabbri}},\ }\bibfield  {title} {\enquote {\bibinfo {title} {Noise
  spectroscopy of a quantum-classical environment with a diamond qubit},}\
  }\href {\doibase 10.1103/PhysRevB.98.214307} {\bibfield  {journal} {\bibinfo
  {journal} {Phys. Rev. B}\ }\textbf {\bibinfo {volume} {98}},\ \bibinfo
  {pages} {214307} (\bibinfo {year} {2018})}\BibitemShut {NoStop}%
\bibitem [{\citenamefont {Childress}\ \emph {et~al.}(2006)\citenamefont
  {Childress}, \citenamefont {{Gurudev Dutt}}, \citenamefont {Taylor},
  \citenamefont {Zibrov}, \citenamefont {Jelezko}, \citenamefont {Wrachtrup},
  \citenamefont {Hemmer},\ and\ \citenamefont {Lukin}}]{Childress_Science06}%
  \BibitemOpen
  \bibfield  {author} {\bibinfo {author} {\bibfnamefont {L.}~\bibnamefont
  {Childress}}, \bibinfo {author} {\bibfnamefont {M.~V.}\ \bibnamefont
  {{Gurudev Dutt}}}, \bibinfo {author} {\bibfnamefont {J.~M.}\ \bibnamefont
  {Taylor}}, \bibinfo {author} {\bibfnamefont {A.~S.}\ \bibnamefont {Zibrov}},
  \bibinfo {author} {\bibfnamefont {F.}~\bibnamefont {Jelezko}}, \bibinfo
  {author} {\bibfnamefont {J.}~\bibnamefont {Wrachtrup}}, \bibinfo {author}
  {\bibfnamefont {P.~R.}\ \bibnamefont {Hemmer}}, \ and\ \bibinfo {author}
  {\bibfnamefont {M.~D.}\ \bibnamefont {Lukin}},\ }\bibfield  {title} {\enquote
  {\bibinfo {title} {Coherent dynamics of coupled electron and nuclear spin
  qubits in diamond},}\ }\href {\doibase 10.1126/science.1131871} {\bibfield
  {journal} {\bibinfo  {journal} {Science}\ }\textbf {\bibinfo {volume}
  {314}},\ \bibinfo {pages} {281} (\bibinfo {year} {2006})}\BibitemShut
  {NoStop}%
\bibitem [{\citenamefont {{Gurudev Dutt}}\ \emph {et~al.}(2007)\citenamefont
  {{Gurudev Dutt}}, \citenamefont {Childress}, \citenamefont {Jiang},
  \citenamefont {Togan}, \citenamefont {Maze}, \citenamefont {Jelezko},
  \citenamefont {Zibrov}, \citenamefont {Hemmer},\ and\ \citenamefont
  {Lukin}}]{Dutt_Science07}%
  \BibitemOpen
  \bibfield  {author} {\bibinfo {author} {\bibfnamefont {M.~V.}\ \bibnamefont
  {{Gurudev Dutt}}}, \bibinfo {author} {\bibfnamefont {L.}~\bibnamefont
  {Childress}}, \bibinfo {author} {\bibfnamefont {L.}~\bibnamefont {Jiang}},
  \bibinfo {author} {\bibfnamefont {E.}~\bibnamefont {Togan}}, \bibinfo
  {author} {\bibfnamefont {J.}~\bibnamefont {Maze}}, \bibinfo {author}
  {\bibfnamefont {F.}~\bibnamefont {Jelezko}}, \bibinfo {author} {\bibfnamefont
  {A.~S.}\ \bibnamefont {Zibrov}}, \bibinfo {author} {\bibfnamefont {P.~R.}\
  \bibnamefont {Hemmer}}, \ and\ \bibinfo {author} {\bibfnamefont {M.~D.}\
  \bibnamefont {Lukin}},\ }\bibfield  {title} {\enquote {\bibinfo {title}
  {Quantum register based on individual electronic and nuclear spin qubits in
  diamond},}\ }\href {\doibase 10.1126/science.1139831} {\bibfield  {journal}
  {\bibinfo  {journal} {Science}\ }\textbf {\bibinfo {volume} {316}},\ \bibinfo
  {pages} {1312} (\bibinfo {year} {2007})}\BibitemShut {NoStop}%
\bibitem [{\citenamefont {Jiang}\ \emph {et~al.}(2009)\citenamefont {Jiang},
  \citenamefont {Hodges}, \citenamefont {Maze}, \citenamefont {Maurer},
  \citenamefont {Taylor}, \citenamefont {Cory}, \citenamefont {Hemmer},
  \citenamefont {Walsworth}, \citenamefont {Yacoby}, \citenamefont {Zibrov},\
  and\ \citenamefont {Lukin}}]{Jiang_Science09}%
  \BibitemOpen
  \bibfield  {author} {\bibinfo {author} {\bibfnamefont {L.}~\bibnamefont
  {Jiang}}, \bibinfo {author} {\bibfnamefont {J.~S.}\ \bibnamefont {Hodges}},
  \bibinfo {author} {\bibfnamefont {J.~R.}\ \bibnamefont {Maze}}, \bibinfo
  {author} {\bibfnamefont {P.}~\bibnamefont {Maurer}}, \bibinfo {author}
  {\bibfnamefont {J.~M.}\ \bibnamefont {Taylor}}, \bibinfo {author}
  {\bibfnamefont {D.~G.}\ \bibnamefont {Cory}}, \bibinfo {author}
  {\bibfnamefont {P.~R.}\ \bibnamefont {Hemmer}}, \bibinfo {author}
  {\bibfnamefont {R.~L.}\ \bibnamefont {Walsworth}}, \bibinfo {author}
  {\bibfnamefont {A.}~\bibnamefont {Yacoby}}, \bibinfo {author} {\bibfnamefont
  {A.~S.}\ \bibnamefont {Zibrov}}, \ and\ \bibinfo {author} {\bibfnamefont
  {M.~D.}\ \bibnamefont {Lukin}},\ }\bibfield  {title} {\enquote {\bibinfo
  {title} {Repetitive readout of a single electronic spin via quantum logic
  with nuclear spin ancillae},}\ }\href {\doibase 10.1126/science.1176496}
  {\bibfield  {journal} {\bibinfo  {journal} {Science}\ }\textbf {\bibinfo
  {volume} {326}},\ \bibinfo {pages} {267} (\bibinfo {year}
  {2009})}\BibitemShut {NoStop}%
\bibitem [{\citenamefont {Robledo}\ \emph {et~al.}(2011)\citenamefont
  {Robledo}, \citenamefont {Childress}, \citenamefont {Bernien}, \citenamefont
  {Hensen}, \citenamefont {Alkemade},\ and\ \citenamefont
  {Hanson}}]{Robledo_Nature11}%
  \BibitemOpen
  \bibfield  {author} {\bibinfo {author} {\bibfnamefont {Lucio}\ \bibnamefont
  {Robledo}}, \bibinfo {author} {\bibfnamefont {Lilian}\ \bibnamefont
  {Childress}}, \bibinfo {author} {\bibfnamefont {Hannes}\ \bibnamefont
  {Bernien}}, \bibinfo {author} {\bibfnamefont {Bas}\ \bibnamefont {Hensen}},
  \bibinfo {author} {\bibfnamefont {Paul F.~A.}\ \bibnamefont {Alkemade}}, \
  and\ \bibinfo {author} {\bibfnamefont {Ronald}\ \bibnamefont {Hanson}},\
  }\bibfield  {title} {\enquote {\bibinfo {title} {High-fidelity projective
  read-out of a solid-state spin quantum register},}\ }\href {\doibase
  10.1038/nature10401} {\bibfield  {journal} {\bibinfo  {journal} {Nature}\
  }\textbf {\bibinfo {volume} {477}},\ \bibinfo {pages} {574} (\bibinfo {year}
  {2011})}\BibitemShut {NoStop}%
\bibitem [{\citenamefont {Taminiau}\ \emph {et~al.}(2014)\citenamefont
  {Taminiau}, \citenamefont {Cramer}, \citenamefont {{van der Sar}},
  \citenamefont {Dobrovitski},\ and\ \citenamefont {Hanson}}]{Taminiau_NN14}%
  \BibitemOpen
  \bibfield  {author} {\bibinfo {author} {\bibfnamefont {T.~H.}\ \bibnamefont
  {Taminiau}}, \bibinfo {author} {\bibfnamefont {J.}~\bibnamefont {Cramer}},
  \bibinfo {author} {\bibfnamefont {T.}~\bibnamefont {{van der Sar}}}, \bibinfo
  {author} {\bibfnamefont {V.~V.}\ \bibnamefont {Dobrovitski}}, \ and\ \bibinfo
  {author} {\bibfnamefont {R.}~\bibnamefont {Hanson}},\ }\bibfield  {title}
  {\enquote {\bibinfo {title} {Universal control and error correction in
  multi-qubit spin registers in diamond},}\ }\href {\doibase
  10.1038/nnano.2014.2} {\bibfield  {journal} {\bibinfo  {journal} {Nature
  Nanotechnology}\ }\textbf {\bibinfo {volume} {9}},\ \bibinfo {pages} {171}
  (\bibinfo {year} {2014})}\BibitemShut {NoStop}%
\bibitem [{\citenamefont {Waldherr}\ \emph {et~al.}(2014)\citenamefont
  {Waldherr}, \citenamefont {Wang}, \citenamefont {Zaiser}, \citenamefont
  {Jamali}, \citenamefont {Schulte-Herbr\"uggen}, \citenamefont {Abe},
  \citenamefont {Ohshima}, \citenamefont {Isoya}, \citenamefont {Du},
  \citenamefont {Neumann},\ and\ \citenamefont
  {Wrachtrup}}]{Waldherr_Nature14}%
  \BibitemOpen
  \bibfield  {author} {\bibinfo {author} {\bibfnamefont {G.}~\bibnamefont
  {Waldherr}}, \bibinfo {author} {\bibfnamefont {Y.}~\bibnamefont {Wang}},
  \bibinfo {author} {\bibfnamefont {S.}~\bibnamefont {Zaiser}}, \bibinfo
  {author} {\bibfnamefont {M.}~\bibnamefont {Jamali}}, \bibinfo {author}
  {\bibfnamefont {T.}~\bibnamefont {Schulte-Herbr\"uggen}}, \bibinfo {author}
  {\bibfnamefont {H.}~\bibnamefont {Abe}}, \bibinfo {author} {\bibfnamefont
  {T.}~\bibnamefont {Ohshima}}, \bibinfo {author} {\bibfnamefont
  {J.}~\bibnamefont {Isoya}}, \bibinfo {author} {\bibfnamefont {J.~F.}\
  \bibnamefont {Du}}, \bibinfo {author} {\bibfnamefont {P.}~\bibnamefont
  {Neumann}}, \ and\ \bibinfo {author} {\bibfnamefont {J.}~\bibnamefont
  {Wrachtrup}},\ }\bibfield  {title} {\enquote {\bibinfo {title} {Quantum error
  correction in a solid-state hybrid spin register},}\ }\href {\doibase
  10.1038/nature12919} {\bibfield  {journal} {\bibinfo  {journal} {Nature}\
  }\textbf {\bibinfo {volume} {506}},\ \bibinfo {pages} {204} (\bibinfo {year}
  {2014})}\BibitemShut {NoStop}%
\bibitem [{\citenamefont {Bradley}\ \emph {et~al.}(2019)\citenamefont
  {Bradley}, \citenamefont {Randall}, \citenamefont {Abobeih}, \citenamefont
  {Berrevoets}, \citenamefont {Degen}, \citenamefont {Bakker}, \citenamefont
  {Markham}, \citenamefont {Twitchen},\ and\ \citenamefont
  {Taminiau}}]{Bradley_PRX19}%
  \BibitemOpen
  \bibfield  {author} {\bibinfo {author} {\bibfnamefont {C.~E.}\ \bibnamefont
  {Bradley}}, \bibinfo {author} {\bibfnamefont {J.}~\bibnamefont {Randall}},
  \bibinfo {author} {\bibfnamefont {M.~H.}\ \bibnamefont {Abobeih}}, \bibinfo
  {author} {\bibfnamefont {R.~C.}\ \bibnamefont {Berrevoets}}, \bibinfo
  {author} {\bibfnamefont {M.~J.}\ \bibnamefont {Degen}}, \bibinfo {author}
  {\bibfnamefont {M.~A.}\ \bibnamefont {Bakker}}, \bibinfo {author}
  {\bibfnamefont {M.}~\bibnamefont {Markham}}, \bibinfo {author} {\bibfnamefont
  {D.~J.}\ \bibnamefont {Twitchen}}, \ and\ \bibinfo {author} {\bibfnamefont
  {T.~H.}\ \bibnamefont {Taminiau}},\ }\bibfield  {title} {\enquote {\bibinfo
  {title} {A ten-qubit solid-state spin register with quantum memory up to one
  minute},}\ }\href {\doibase 10.1103/PhysRevX.9.031045} {\bibfield  {journal}
  {\bibinfo  {journal} {Phys. Rev. X}\ }\textbf {\bibinfo {volume} {9}},\
  \bibinfo {pages} {031045} (\bibinfo {year} {2019})}\BibitemShut {NoStop}%
\bibitem [{\citenamefont {Zhao}\ \emph {et~al.}(2011)\citenamefont {Zhao},
  \citenamefont {Wang},\ and\ \citenamefont {Liu}}]{Zhao_PRL11}%
  \BibitemOpen
  \bibfield  {author} {\bibinfo {author} {\bibfnamefont {Nan}\ \bibnamefont
  {Zhao}}, \bibinfo {author} {\bibfnamefont {Zhen-Yu}\ \bibnamefont {Wang}}, \
  and\ \bibinfo {author} {\bibfnamefont {Ren-Bao}\ \bibnamefont {Liu}},\
  }\bibfield  {title} {\enquote {\bibinfo {title} {Anomalous decoherence effect
  in a quantum bath},}\ }\href {\doibase 10.1103/PhysRevLett.106.217205}
  {\bibfield  {journal} {\bibinfo  {journal} {Phys.\ Rev.\ Lett.}\ }\textbf
  {\bibinfo {volume} {106}},\ \bibinfo {pages} {217205} (\bibinfo {year}
  {2011})}\BibitemShut {NoStop}%
\bibitem [{\citenamefont {Lang}\ \emph {et~al.}(2019)\citenamefont {Lang},
  \citenamefont {Madhavan}, \citenamefont {Tetienne}, \citenamefont {Broadway},
  \citenamefont {Hall}, \citenamefont {Teraji}, \citenamefont {Monteiro},
  \citenamefont {Stacey},\ and\ \citenamefont {Hollenberg}}]{Lang_PRA19}%
  \BibitemOpen
  \bibfield  {author} {\bibinfo {author} {\bibfnamefont {J.~E.}\ \bibnamefont
  {Lang}}, \bibinfo {author} {\bibfnamefont {T.}~\bibnamefont {Madhavan}},
  \bibinfo {author} {\bibfnamefont {J.-P.}\ \bibnamefont {Tetienne}}, \bibinfo
  {author} {\bibfnamefont {D.~A.}\ \bibnamefont {Broadway}}, \bibinfo {author}
  {\bibfnamefont {L.~T.}\ \bibnamefont {Hall}}, \bibinfo {author}
  {\bibfnamefont {T.}~\bibnamefont {Teraji}}, \bibinfo {author} {\bibfnamefont
  {T.~S.}\ \bibnamefont {Monteiro}}, \bibinfo {author} {\bibfnamefont
  {A.}~\bibnamefont {Stacey}}, \ and\ \bibinfo {author} {\bibfnamefont
  {L.~C.~L.}\ \bibnamefont {Hollenberg}},\ }\bibfield  {title} {\enquote
  {\bibinfo {title} {Nonvanishing effect of detuning errors in
  dynamical-decoupling-based quantum sensing experiments},}\ }\href {\doibase
  10.1103/PhysRevA.99.012110} {\bibfield  {journal} {\bibinfo  {journal} {Phys.
  Rev. A}\ }\textbf {\bibinfo {volume} {99}},\ \bibinfo {pages} {012110}
  (\bibinfo {year} {2019})}\BibitemShut {NoStop}%
\bibitem [{\citenamefont {Cywi{\'n}ski}(2014)}]{Cywinski_PRA14}%
  \BibitemOpen
  \bibfield  {author} {\bibinfo {author} {\bibfnamefont {{\L}ukasz}\
  \bibnamefont {Cywi{\'n}ski}},\ }\bibfield  {title} {\enquote {\bibinfo
  {title} {Dynamical-decoupling noise spectroscopy at an optimal working point
  of a qubit},}\ }\href {\doibase 10.1103/PhysRevA.90.042307} {\bibfield
  {journal} {\bibinfo  {journal} {Phys. Rev. A}\ }\textbf {\bibinfo {volume}
  {90}},\ \bibinfo {pages} {042307} (\bibinfo {year} {2014})}\BibitemShut
  {NoStop}%
\bibitem [{\citenamefont {Sung}\ \emph {et~al.}(2019)\citenamefont {Sung},
  \citenamefont {Beaudoin}, \citenamefont {Norris}, \citenamefont {Yan},
  \citenamefont {Kim}, \citenamefont {Qiu}, \citenamefont {{von L\"uepke}},
  \citenamefont {Yoder}, \citenamefont {Orlando}, \citenamefont {Viola},
  \citenamefont {Gustavsson},\ and\ \citenamefont {Oliver}}]{Sung_NC19}%
  \BibitemOpen
  \bibfield  {author} {\bibinfo {author} {\bibfnamefont {Youngkyu}\
  \bibnamefont {Sung}}, \bibinfo {author} {\bibfnamefont {F\'elix}\
  \bibnamefont {Beaudoin}}, \bibinfo {author} {\bibfnamefont {Leigh~M.}\
  \bibnamefont {Norris}}, \bibinfo {author} {\bibfnamefont {Fei}\ \bibnamefont
  {Yan}}, \bibinfo {author} {\bibfnamefont {David~K.}\ \bibnamefont {Kim}},
  \bibinfo {author} {\bibfnamefont {Jack~Y.}\ \bibnamefont {Qiu}}, \bibinfo
  {author} {\bibfnamefont {Uwe}\ \bibnamefont {{von L\"uepke}}}, \bibinfo
  {author} {\bibfnamefont {Jonilyn~L.}\ \bibnamefont {Yoder}}, \bibinfo
  {author} {\bibfnamefont {Terry~P.}\ \bibnamefont {Orlando}}, \bibinfo
  {author} {\bibfnamefont {Lorenza}\ \bibnamefont {Viola}}, \bibinfo {author}
  {\bibfnamefont {Simon}\ \bibnamefont {Gustavsson}}, \ and\ \bibinfo {author}
  {\bibfnamefont {William~D.}\ \bibnamefont {Oliver}},\ }\bibfield  {title}
  {\enquote {\bibinfo {title} {Non-gaussian noise spectroscopy with a
  superconducting qubit sensor},}\ }\href {\doibase 10.1038/s41467-019-11699-4}
  {\bibfield  {journal} {\bibinfo  {journal} {Nat.~Commun.}\ }\textbf {\bibinfo
  {volume} {10}},\ \bibinfo {pages} {3715} (\bibinfo {year}
  {2019})}\BibitemShut {NoStop}%
\bibitem [{\citenamefont {Norris}\ \emph {et~al.}(2016)\citenamefont {Norris},
  \citenamefont {Paz-Silva},\ and\ \citenamefont {Viola}}]{Norris_PRL16}%
  \BibitemOpen
  \bibfield  {author} {\bibinfo {author} {\bibfnamefont {Leigh~M.}\
  \bibnamefont {Norris}}, \bibinfo {author} {\bibfnamefont {Gerardo~A.}\
  \bibnamefont {Paz-Silva}}, \ and\ \bibinfo {author} {\bibfnamefont {Lorenza}\
  \bibnamefont {Viola}},\ }\bibfield  {title} {\enquote {\bibinfo {title}
  {Qubit noise spectroscopy for non-gaussian dephasing environments},}\ }\href
  {\doibase 10.1103/PhysRevLett.116.150503} {\bibfield  {journal} {\bibinfo
  {journal} {Phys. Rev. Lett.}\ }\textbf {\bibinfo {volume} {116}},\ \bibinfo
  {pages} {150503} (\bibinfo {year} {2016})}\BibitemShut {NoStop}%
\bibitem [{\citenamefont {Roszak}\ \emph {et~al.}(2019)\citenamefont {Roszak},
  \citenamefont {Kwiatkowski},\ and\ \citenamefont
  {Cywi\'{n}ski}}]{Roszak_PRA19}%
  \BibitemOpen
  \bibfield  {author} {\bibinfo {author} {\bibfnamefont {Katarzyna}\
  \bibnamefont {Roszak}}, \bibinfo {author} {\bibfnamefont {Damian}\
  \bibnamefont {Kwiatkowski}}, \ and\ \bibinfo {author} {\bibfnamefont
  {\L{}ukasz}\ \bibnamefont {Cywi\'{n}ski}},\ }\bibfield  {title} {\enquote
  {\bibinfo {title} {How to detect qubit-environment entanglement generated
  during qubit dephasing},}\ }\href {\doibase 10.1103/PhysRevA.100.022318}
  {\bibfield  {journal} {\bibinfo  {journal} {Phys. Rev. A}\ }\textbf {\bibinfo
  {volume} {100}},\ \bibinfo {pages} {022318} (\bibinfo {year}
  {2019})}\BibitemShut {NoStop}%
\bibitem [{\citenamefont {Krummheuer}\ \emph {et~al.}(2002)\citenamefont
  {Krummheuer}, \citenamefont {Axt},\ and\ \citenamefont
  {Kuhn}}]{Krummheuer_PRB02}%
  \BibitemOpen
  \bibfield  {author} {\bibinfo {author} {\bibfnamefont {B.}~\bibnamefont
  {Krummheuer}}, \bibinfo {author} {\bibfnamefont {V.~M.}\ \bibnamefont {Axt}},
  \ and\ \bibinfo {author} {\bibfnamefont {T.}~\bibnamefont {Kuhn}},\
  }\bibfield  {title} {\enquote {\bibinfo {title} {Theory of pure dephasing and
  the resulting absorption line shape in semiconductor quantum dots},}\ }\href
  {\doibase 10.1103/PhysRevB.65.195313} {\bibfield  {journal} {\bibinfo
  {journal} {Phys.\ Rev.\ B}\ }\textbf {\bibinfo {volume} {65}},\ \bibinfo
  {pages} {195313} (\bibinfo {year} {2002})}\BibitemShut {NoStop}%
\bibitem [{\citenamefont {Roszak}\ and\ \citenamefont
  {Machnikowski}(2006{\natexlab{a}})}]{Roszak_PLA06}%
  \BibitemOpen
  \bibfield  {author} {\bibinfo {author} {\bibfnamefont {K.}~\bibnamefont
  {Roszak}}\ and\ \bibinfo {author} {\bibfnamefont {P.}~\bibnamefont
  {Machnikowski}},\ }\bibfield  {title} {\enquote {\bibinfo {title} {{``Which
  path''} decoherence in quantum dot experiments},}\ }\href {\doibase
  10.1016/j.physleta.2005.11.012} {\bibfield  {journal} {\bibinfo  {journal}
  {Phys.~Lett.~A}\ }\textbf {\bibinfo {volume} {351}},\ \bibinfo {pages}
  {251--256} (\bibinfo {year} {2006}{\natexlab{a}})}\BibitemShut {NoStop}%
\bibitem [{\citenamefont {Roszak}\ and\ \citenamefont
  {Machnikowski}(2006{\natexlab{b}})}]{Roszak_PRA06}%
  \BibitemOpen
  \bibfield  {author} {\bibinfo {author} {\bibfnamefont {Katarzyna}\
  \bibnamefont {Roszak}}\ and\ \bibinfo {author} {\bibfnamefont {Pawe\l{}}\
  \bibnamefont {Machnikowski}},\ }\bibfield  {title} {\enquote {\bibinfo
  {title} {Complete disentanglement by partial pure dephasing},}\ }\href
  {\doibase 10.1103/PhysRevA.73.022313} {\bibfield  {journal} {\bibinfo
  {journal} {Phys. Rev. A}\ }\textbf {\bibinfo {volume} {73}},\ \bibinfo
  {pages} {022313} (\bibinfo {year} {2006}{\natexlab{b}})}\BibitemShut
  {NoStop}%
\bibitem [{\citenamefont {Hodgson}\ \emph {et~al.}(2008)\citenamefont
  {Hodgson}, \citenamefont {Viola},\ and\ \citenamefont
  {D'Amico}}]{Hodgson_PRB08}%
  \BibitemOpen
  \bibfield  {author} {\bibinfo {author} {\bibfnamefont {Thomas~E.}\
  \bibnamefont {Hodgson}}, \bibinfo {author} {\bibfnamefont {Lorenza}\
  \bibnamefont {Viola}}, \ and\ \bibinfo {author} {\bibfnamefont {Irene}\
  \bibnamefont {D'Amico}},\ }\bibfield  {title} {\enquote {\bibinfo {title}
  {Decoherence-protected storage of exciton qubits through ultrafast multipulse
  control},}\ }\href {\doibase 10.1103/PhysRevB.78.165311} {\bibfield
  {journal} {\bibinfo  {journal} {Phys. Rev. B}\ }\textbf {\bibinfo {volume}
  {78}},\ \bibinfo {pages} {165311} (\bibinfo {year} {2008})}\BibitemShut
  {NoStop}%
\bibitem [{\citenamefont {Ramon}(2012)}]{Ramon_PRB12}%
  \BibitemOpen
  \bibfield  {author} {\bibinfo {author} {\bibfnamefont {Guy}\ \bibnamefont
  {Ramon}},\ }\bibfield  {title} {\enquote {\bibinfo {title} {Dynamical
  decoupling of a singlet-triplet qubit afflicted by a charge fluctuator},}\
  }\href {\doibase 10.1103/PhysRevB.86.125317} {\bibfield  {journal} {\bibinfo
  {journal} {Phys.\ Rev.\ B}\ }\textbf {\bibinfo {volume} {86}},\ \bibinfo
  {pages} {125317} (\bibinfo {year} {2012})}\BibitemShut {NoStop}%
\bibitem [{\citenamefont {Shulman}\ \emph {et~al.}(2012)\citenamefont
  {Shulman}, \citenamefont {Dial}, \citenamefont {Harvey}, \citenamefont
  {Bluhm}, \citenamefont {Umansky},\ and\ \citenamefont
  {Yacoby}}]{Shulman_Science12}%
  \BibitemOpen
  \bibfield  {author} {\bibinfo {author} {\bibfnamefont {M.~D.}\ \bibnamefont
  {Shulman}}, \bibinfo {author} {\bibfnamefont {O.~E.}\ \bibnamefont {Dial}},
  \bibinfo {author} {\bibfnamefont {S.~P.}\ \bibnamefont {Harvey}}, \bibinfo
  {author} {\bibfnamefont {H.}~\bibnamefont {Bluhm}}, \bibinfo {author}
  {\bibfnamefont {V.}~\bibnamefont {Umansky}}, \ and\ \bibinfo {author}
  {\bibfnamefont {A.}~\bibnamefont {Yacoby}},\ }\bibfield  {title} {\enquote
  {\bibinfo {title} {Demonstration of entanglement of electrostatically coupled
  singlet-triplet qubits},}\ }\href {\doibase 10.1126/science.1217692}
  {\bibfield  {journal} {\bibinfo  {journal} {Science}\ }\textbf {\bibinfo
  {volume} {336}},\ \bibinfo {pages} {202} (\bibinfo {year}
  {2012})}\BibitemShut {NoStop}%
\bibitem [{\citenamefont {Ramon}(2015)}]{Ramon_PRB15}%
  \BibitemOpen
  \bibfield  {author} {\bibinfo {author} {\bibfnamefont {Guy}\ \bibnamefont
  {Ramon}},\ }\bibfield  {title} {\enquote {\bibinfo {title} {Non-gaussian
  signatures and collective effects in charge noise affecting a dynamically
  decoupled qubit},}\ }\href {\doibase 10.1103/PhysRevB.92.155422} {\bibfield
  {journal} {\bibinfo  {journal} {Phys. Rev. B}\ }\textbf {\bibinfo {volume}
  {92}},\ \bibinfo {pages} {155422} (\bibinfo {year} {2015})}\BibitemShut
  {NoStop}%
\end{thebibliography}

%

\end{document}